\documentclass[aps,prl,reprint,superscriptaddress,
 amsmath,amssymb,floatfix]{revtex4-2}
\usepackage{amssymb}
\usepackage{amsmath}
\usepackage[latin9]{inputenc}
\usepackage{array}
\usepackage{multirow}
\usepackage{color}
\usepackage{esint}
\usepackage{bm}
\usepackage{color}
\usepackage{bbm}
\usepackage[breaklinks,colorlinks,
linkcolor=blue,citecolor=blue,urlcolor=blue]{hyperref}
\usepackage{babel}
\usepackage{titlesec}
\usepackage{graphicx,amssymb, color}
\usepackage{epsfig}
\usepackage{epstopdf}
\usepackage{mathtools}
\usepackage{soul}
\usepackage[dvipsnames]{xcolor}
\usepackage{tikz}
\usepackage{tcolorbox}
\usepackage{graphics,amssymb,amsmath,epsfig,color}
\usepackage{simpler-wick}
\usepackage{physics}

\begin{document}
\title{
Dynamical Phase Transition of Dissipative Fermionic Superfluids
}
\author{Xin-Yuan Gao}
\affiliation{%
Department of Physics, The Chinese University of Hong Kong, Shatin, New Territories, Hong Kong, China
}%

\author{Yangqian Yan}%
 \email{yqyan@cuhk.edu.hk}
\affiliation{%
Department of Physics, The Chinese University of Hong Kong, Shatin, New Territories, Hong Kong, China
}
\affiliation{State Key Laboratory of Quantum Information Technologies and Materials, The Chinese University of Hong Kong, Hong Kong SAR, China}
\affiliation{
The Chinese University of Hong Kong Shenzhen Research Institute, 518057 Shenzhen, China
}%
\begin{abstract}
Driven-dissipative open quantum many-body systems exhibit rich phases that are characterized by the steady states in the long-time dynamics.
However, lossy open systems inevitably decay to the vacuum, making their transient evolution the primary focus.
Assuming the Hartree-Fock-Bogoliubov ansatz, we derive a generalized time-dependent Hartree-Fock-Bogoliubov equation based on the least action principle for the open quantum systems.
By solving the quench dynamics after abruptly introducing inelastic scattering or one-body loss in the Bardeen-Cooper-Schrieffer limit, we reveal a generic dynamical phase transition: the superfluid order parameter vanishes non-analytically while the superfluid fraction's first-order time derivative undergoes a discontinuous change at a finite critical time. This marks a new paradigm of dynamical phase transitions, distinct from those in closed systems, where the initial state must be finely tuned.
\end{abstract}
\maketitle

\textit{Introduction.---}
The interplay between coherent evolution and dissipation gives rise to rich physical phenomena in driven-dissipative open quantum many-body systems.
Much attention has been devoted to understanding non-equilibrium phases of different steady states~\cite{kessler2012dissipative,honing2012critical,honing2013steadystate,horstmann2013noisedriven,torre2013keldysh,sieberer2013dynamical,qian2013quantum,leboite2013steadystate,lee2013unconventional,carr2013nonequilibriuma,joshi2013quantum,malossi2014fulla,marcuzzi2014universala,lang2015exploring,overbeck2017multicritical,savona2017spontaneous,tomita2017observation,gao2019steadystate,huybrechts2020validity,jen2020steadystate,landa2020correlationinduced,li2021steadystate}, with numerous theoretical methods developed for their description, including Keldysh field theory \cite{sieberer2016keldysh}, variational approaches \cite{weimer2015variational}, corner-space renormalization \cite{finazzi2015cornerspace}, cluster mean-field theory \cite{jin2016cluster}, and tensor network algorithms \cite{kshetrimayum2017simplea}.
However, purely lossy systems, such as ultracold atomic gases subject to losses~\cite{ni2008high,pagano2015strongly,cappellini2019coherenta,takasu2020ptsymmetric,wang2021photoexcitation,jager2024methods,jager2024precise}, inevitably evolve towards the vacuum state, making steady-state analyses insufficient.
For such systems, the transient dynamics preceding vacuum decay become the primary focus, demanding theoretical frameworks that can capture the full temporal evolution.

\begin{figure}
    \centering
    \includegraphics[width=0.499\textwidth]{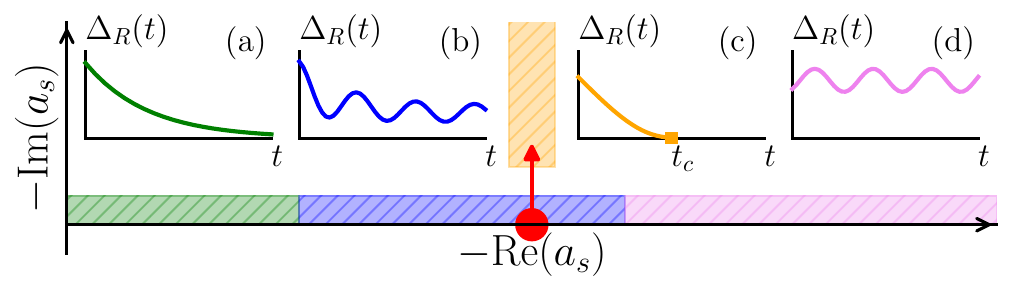}
    \caption{Schematics of quenching the complex scattering length $a_s$ of fermionic superfluids.
Panels show the order parameter as a function of time for different quench schemes.
Panels (a), (b), and (d) are for quench along the real axis~\cite{barankov2004collective,barankov2006synchronization,yuzbashyan2006dynamical,young2024observing} (green, blue, and pink lines).
Panel (c) is for quenching to a complex scattering length, and the red dot denotes the initial state; the order parameter vanishes non-analytically at critical time $t_c$ (square).
}
    \label{fig1}
\end{figure}

A particularly important example is dissipative fermionic superfluids, where inelastic scattering introduces complex dynamics that fundamentally differ from their closed-system counterparts.
While the dynamics of elastic scattering length quenches in superfluids are well-understood, producing oscillating or exponentially decaying order parameters [Fig.~\ref{fig1}(a,b,d)]~\cite{barankov2004collective,barankov2006synchronization,yuzbashyan2006dynamical,young2024observing}, the effects of inelastic scattering have received limited theoretical attention.
Previous studies using Anderson's pseudo-spin formalism on lattices~\cite{yamamoto2019theory,yamamoto2021collective,mazza2023dissipative} assume a pure superfluid without accounting for the normal component that emerges during dissipative evolution.
This gap motivates our investigation of homogeneous lossy fermionic superfluids in continuous space, where we discover a fundamentally new type of dynamical phase transition.

In this Letter, we extend the action in closed systems~\cite{blaizot1986quantum} to open quantum systems governed by the Lindblad master equation and derive a generalized time-dependent Hartree-Fock-Bogoliubov equation that captures the evolution of both superfluid and normal components in lossy two-component Fermi gases.
Using semi-classical approximations, we solve this equation analytically for quench dynamics in homogeneous systems within the Bardeen-Cooper-Schrieffer (BCS) limit, where dissipation is suddenly introduced through inelastic scattering or one-body loss.

Our key discovery is a universal dissipative dynamical phase transition that occurs at a finite critical time $t_c$, where the system transitions from superfluid to normal fluid.
This transition is featured by discontinuous first-order time derivatives in both quasi-particle distribution and superfluid fraction, and non-analytic decay of the order parameter.
Crucially, unlike dynamical phase transitions in closed systems that require finely tuned initial states~\cite{heyl2013dynamical,heyl2018dynamical}, our dissipative transition occurs universally for any BCS ground state, regardless of the initial elastic scattering length.
This universality represents a new paradigm for dynamical phase transitions, where dissipation itself drives the critical behavior rather than specific initial state preparation.

\textit{Hamiltonian.---} We consider a spin-balanced Fermi gas under a general external potential $U$. The momentum-space Hamiltonian reads
\begin{align}
    \widehat{H}\equiv\widehat{H}_R&=\sum_{\mathbf{q},\mathbf{q}',\sigma}[(\epsilon_\mathbf{q}-\mu)\delta_{\mathbf{q},\mathbf{q}'}+U(|\mathbf{q}-\mathbf{q}'|)] c^\dagger_{\mathbf{q},\sigma}c_{\mathbf{q}',\sigma}\nonumber\\
    &+\frac{g_R}{V}\sum_{\mathbf{P},\mathbf{q},\mathbf{q}'}c^\dagger_{\frac{\mathbf{P}}{2}-\mathbf{q}',\uparrow}c^\dagger_{\frac{\mathbf{P}}{2}+\mathbf{q}',\downarrow}c_{\frac{\mathbf{P}}{2}+\mathbf{q},\downarrow}c_{\frac{\mathbf{P}}{2}-\mathbf{q},\uparrow},
    \label{H_R}
\end{align}
where $\mu$, $V$, and $c_{\mathbf{q},\sigma}$ are the chemical potential, system volume, and annihilation operator of a fermion with momentum $\mathbf{q}$ and spin $\sigma$, respectively.
The coupling constant $g_R$ requires proper renormalization
$
    \frac{m}{4 \pi \hbar^2 \mathrm{Re}(a_s)}=\frac{1}{g_R}+ \int \frac{d^3 p}{(2 \pi \hbar)^3}\frac{1}{p^2 / m},
$
where $m$ is the particle mass, and $a_s$ is the $s$-wave scattering length.
We consider two possible channels for dissipation,
\begin{equation}
    \widehat{L}_{1,\mathbf{k},\sigma}=\sqrt{2\gamma}c_{\mathbf{k},\sigma},\  \widehat{L}_{2,\mathbf{P}}=\sqrt{\frac{2g_I}{V}}\sum_\mathbf{q} c_{\frac{\mathbf{P}}{2}+\mathbf{q},\downarrow} c_{\frac{\mathbf{P}}{2}-\mathbf{q},\uparrow},
\end{equation}
modeling one- and two-body losses, respectively.
Here, $\gamma$ is the one-body dissipation rate, and $g_I={4\pi\hbar\mathrm{Im}(a_s)}/{m}$ is the two-body dissipation coupling constant~\cite{wang2022complex}.
The imaginary part of the non-Hermitian Hamiltonian reads $\widehat{H}_I\equiv -\sum_n \widehat{L}^\dagger_n \widehat{L}_n/2$.
As shown in Fig.~\ref{fig1}, we prepare the system to equilibrate under elastic interaction, then at $t=0$, a dissipation is turned on to induce dynamics.

\textit{Time-dependent Hartree-Fock-Bogoliubov theory.---} 
Hartree-Fock-Bogoliubov theory has been derived from an auxiliary observable variational principle in closed systems in Ref.~\cite{blaizot1986quantum}. Below, we generalize the action $S$, 

\begin{equation}
\begin{aligned}
    &S=\hbar\Tr\widehat{D}(t_f)\widehat{A}_f-\int_{t_i}^{t_f}dt\Tr\widehat{A} \bigg\{\hbar \frac{d \widehat{D}}{dt}\\
    &+i[\widehat{H},\widehat{D}]-\hbar\sum_n\left(\frac{1}{2}\{\widehat{L}^\dagger_n\widehat{L}_n,\widehat{D}\}-\widehat{L}_n\widehat{D}\widehat{L}^\dagger_n\right)\bigg\},
    \label{S}
\end{aligned}
\end{equation}
to open quantum systems and largely follow the derivation;
Here, $\widehat{H}$ is the Hamiltonian, $\widehat{L}_n$ represents the jump operator in channel $n$, and $\Tr \widehat{D} \widehat{A}$ is the trace of an auxiliary observable $\widehat{A}$ to the density matrix $\widehat{D}$.
The generalization is motivated by replacing the von Neumann equation-like structure in the action by the Lindblad master equation, i.e., we have only added one extra term involving the dissipators, and everything else remains unchanged.
As a sanity check, we verify that the action reproduces the Lindblad equation and adjoint equation (See End Matter).

The advantage of using the action formulation over directly working with the Lindblad equation is that it naturally accommodates constraints through variational principles.
To implement the Hartree-Fock-Bogoliubov approximation, we adopt a fermionic Gaussian ansatz for the density matrix $\widehat{D}_\mathrm{a}(t)=e^{\widehat{W}(t)}/\mathrm{Tr}\left[e^{\widehat{W}(t)}\right]$ and consider quadratic observables $\widehat{A}_\mathrm{a}(t)$, where $\widehat{W}(t)$ and $\widehat{A}_\mathrm{a}(t)$ have general Hermitian quadratic forms (see End Matter and Ref.~\cite{blaizot1986quantum}). This Gaussian ansatz enables the application of generalized Wick's theorem for evaluating expectation values~\cite{blaizot1986quantum}.

Within this framework, we define generalized one-body density matrix $\mathcal{R}$ and observable $\mathcal{A}$ in the single-particle space:
\begin{equation}
    \mathcal{R}=\left(\begin{matrix}
        \rho & \kappa\\
        \kappa^\dagger &1-\rho^*
    \end{matrix}\right),
    \quad
    \mathcal{A}=\left(\begin{matrix}
        a & b\\
        b^\dagger &-a^*
    \end{matrix}\right),
\end{equation}
where $\rho$, $\kappa$, $a$, and $b$ are matrices with momentum indices $\mathbf{k}$ and $\mathbf{k}'$. For spin-balanced systems with inter-species interactions, we consider $\rho_{\mathbf{k},\mathbf{k}'}\equiv\rho_{\mathbf{k},\uparrow,\mathbf{k}',\uparrow}=\rho_{\mathbf{k},\downarrow,\mathbf{k}',\downarrow}$ and $\kappa_{\mathbf{k},\mathbf{k}'}\equiv\kappa_{\mathbf{k},\downarrow,\mathbf{k}',\uparrow}=\kappa_{\mathbf{k},\uparrow,\mathbf{k}',\downarrow}$, where $\rho_{\mathbf{k},\sigma,\mathbf{k}',\sigma'}=\langle c^\dagger_{\mathbf{k'},\sigma'} c_{\mathbf{k},\sigma}\rangle$ is the normal one-body density matrix and $\kappa_{\mathbf{k},\sigma,\mathbf{k}',\sigma'}=\langle c_{\mathbf{k'},\sigma'} c_{\mathbf{k},\sigma}\rangle$ is the anomalous one-body density matrix (pair coherence), with $\langle\cdot\rangle=\operatorname{Tr}(\widehat{D}_\mathrm{a}(t) \cdot)$. Similarly, we denote the coefficients of the quadratic operators in $\widehat{A}_a$ as $a_{\mathbf{k},\mathbf{k}'}\equiv a_{\mathbf{k},\uparrow,\mathbf{k}',\uparrow}=a_{\mathbf{k},\downarrow,\mathbf{k}',\downarrow}$ and $b_{\mathbf{k},\mathbf{k}'}\equiv b_{\mathbf{k},\downarrow,\mathbf{k}',\uparrow}=b_{\mathbf{k},\uparrow,\mathbf{k}',\downarrow}$ (see End Matter).

Applying the Hartree-Fock-Bogoliubov approximation to the action in Eq.~(\ref{S}), we obtain the effective action in terms of $\mathcal{R}$ and $\mathcal{A}$~\cite{blaizot1986quantum,SM}:
\begin{equation}
\begin{aligned}
    S[\mathcal{R},\mathcal{A}]&=\tr\hbar\mathcal{R}(t_f)\mathcal{A}(t_f)-\int_{t_i}^{t_f}dt\tr\hbar\mathcal{A}\big\{\frac{d\mathcal{R}}{dt}\\
    &+i[\mathcal{H}_R[\mathcal{R}],\mathcal{R}]-\hbar(\{\mathcal{H}_I[\mathcal{R}],\mathcal{R}\}+\mathcal{J}[\mathcal{R}])\big\},
\end{aligned}
\label{Seff}
\end{equation}
where $\tr$ denotes tracing over the single-particle space,
$\mathcal{H}_R+i\hbar \mathcal{H}_I$ is the effective non-Hermitian Hamiltonian,
\begin{equation}
    \mathcal{H}_{R/I}=\left(\begin{matrix}
        h_{R/I} & \Delta_{R/I}\\
        \Delta_{R/I}^\dagger & -{h}^*_{R/I} 
    \end{matrix}\right),
\end{equation}
with $h_{R/I}$ and $\Delta_{R/I}$ being real (imaginary) Hartree-Fock Hamiltonians and pairing fields, respectively.
They are related to the original Hamiltonian $\widehat{H}$ by
\begin{equation}
    (h_{R/I})_{\mathbf{k},\mathbf{k}'}=\frac{\partial \langle\widehat{H}_{R/I}\rangle}{\partial \rho_{\mathbf{k}',\uparrow,\mathbf{k},\uparrow}},~ (\Delta_{R/I})_{\mathbf{k},\mathbf{k}'}=\frac{\partial  \langle \widehat{H}_{R/I}\rangle}{\partial \kappa^*_{\mathbf{k},\uparrow,\mathbf{k}',\downarrow}}.
\label{variation}
\end{equation}
Besides the non-Hermitian generalization $\mathcal{H}_{R}+i \hbar \mathcal{H}_{I}$ to the Hermitian Hamiltonian $\mathcal{H}_{R}$, we find an extra term, i.e., the effective quantum jump $\mathcal{J}$,
\begin{equation}
    \mathcal{J}=\left(\begin{matrix}
    0 & 2\kappa {h}^*_I-2\rho\Delta_I\\
    2{h}^*_I\kappa^\dagger-2\Delta_I^\dagger\rho & \{{h}^*_I,1-2{\rho^*}\}-2\Delta_I^\dagger\kappa-2\kappa^\dagger\Delta_I 
    \end{matrix}\right).
\end{equation}

Looking for stationary action with respect to the observable $\mathcal{A}$, $\delta S[\mathcal{R},\mathcal{A}]/\delta \mathcal{A}=0$, we obtain the effective equation of motion for the generalized density matrix $\mathcal{R}$:
\begin{equation}
i\hbar \dot{\mathcal{R}}=[\mathcal{H}_R,\mathcal{R}]+i\hbar\{\mathcal{H}_I,\mathcal{R}\}+i\hbar\mathcal{J}.
\label{TDHFB}
\end{equation}
Equation~(\ref{TDHFB}) with $\mathcal{H}_I=0$ and $\mathcal{J}=0$ is known as the time-dependent Hartree-Fock-Bogoliubov equation~\cite{blaizot1986quantum} in closed systems, which has been applied to study giant resonances in nuclear physics~\cite{peter1980nuclear} and collective modes in trapped atomic fermionic superfluids~\cite{urban2006dynamics}. This generalization to open systems is one of the main results of this work. It is worth noting that Eq.~(\ref{TDHFB}) is exact when only one-body loss is present and the interaction is turned off, because it has been proved that for a quadratic Hamiltonian and linear dissipators in fermionic variables, the density matrix remains Gaussian~\cite{bravyi2005lagrangian,prosen2008third}. 

Note that Eq.~(\ref{TDHFB}) does not couple with $\mathcal{A}$, which indicates its generality for applying to situations with arbitrary $t_f$ and $\widehat{A}_f$ that can be written in the form of Eq.~(\ref{Aa}).
Thus, we do not need to further supply with $\delta S[\mathcal{R},\mathcal{A}]/\delta\mathcal{R}=0$ (explicit equation of motion is shown in the End Matter). 

For a system in the normal phase without pairing coherence, e.g., at high temperatures, the off-diagonal terms are zero. Since the left upper block of $\mathcal{J}$ in Eq.~(\ref{TDHFB}) is zero, the equation of motion [Eq.~(\ref{TDHFB})] reduces to
\begin{equation}
    i\hbar \dot{\rho}=[{h}_R,\rho]+i\hbar\{{h}_I,\rho\},
    \label{TDHF}
\end{equation}
which resembles the von Neumann equation for the density matrix $\widehat{D}$ with a non-Hermitian Hamiltonian.

For practical calculations in atomic gases with a large number of particles, we employ a standard semi-classical approximation using the phase-space formulation of quantum mechanics~\cite{dalton2014phase}.
Specifically, we perform a Wigner transform on all quantities in Eq.~(\ref{TDHFB}) (see Supplemental Material~\cite{SM}), e.g., for the pair coherence,
$
    \kappa(\mathbf{r}, \mathbf{p})=\int \frac{d^3 q}{(2\pi\hbar)^3} e^{i \mathbf{q} \cdot \mathbf{r} / \hbar} \kappa_{(\mathbf{p}+\mathbf{q}/2)/\hbar,(\mathbf{p}-\mathbf{q}/2)/\hbar}.
$
After the transformation, the Wigner transformed gap $\Delta_{I/R}(\mathbf{r})$ reads
\begin{equation}
    \Delta_{I/R}(\mathbf{r})=-g_{I/R} \int \frac{d^3 p'}{(2 \pi \hbar)^3}\kappa(\mathbf{r}, \mathbf{p}').
\label{gap_eqn}
\end{equation}
Similar to the BCS theory, we have the freedom to choose a phase factor $\phi(\mathbf{r})$ at each local point $\mathbf{r}$, thus we can always choose $\phi(\mathbf{r})$ to make $\Delta_R$ and $\Delta_I$ real simultaneously.
Under this gauge, the corresponding transformed Hartree-Fock
Hamiltonians $h_R$ and $h_I$ are
\begin{equation}
\begin{aligned}
    &h_R(\mathbf{r},\mathbf{p})=\frac{[p-\hbar\nabla \phi(\mathbf{r})]^2}{2m}+U(\mathbf{r})-\mu+\dot{\phi}(\mathbf{r})+g_R n(\mathbf{r}),\\
    &h_I(\mathbf{r},\mathbf{p})=-\gamma+g_I n(\mathbf{r}),
\end{aligned}
\end{equation}
where $n(\mathbf{r})=\int d^3p/(2\pi\hbar)^3 \rho(\mathbf{r},\mathbf{p})$ is the local density.
Then, we expand the Wigner transformed Eq.~(\ref{TDHFB}) to first order in $\hbar$ (see Supplemental Material~\cite{SM}), and obtain the semi-classical approximation of Eq.~(\ref{TDHFB}), which yields a Boltzmann-like equation for the equation of motion for Bogoliubov quasi-particle distribution $\nu(\mathbf{r},\mathbf{p})$:
\begin{equation}
    \dot{\nu}=\left\{E,\nu\right\}_c+h_I\left(2\nu-1+{h_R^\mathrm{ev}}/{E^\mathrm{ev}}\right),\label{nu_IQBE}
\end{equation}
where $\{\cdot,\cdot\}_c$ denotes the classical Poisson bracket and all quantities are phase-space distributions, i.e., depends on $\mathbf{r},\mathbf{p}$; the super script ${\mathrm{ev}/\mathrm{od}}$ refers to the time-even/odd parts, i.e., $\nu^{\mathrm{ev}}=[\nu(\mathbf{r},\mathbf{p})+\nu(\mathbf{r},-\mathbf{p})]/2$ and $\nu^{\mathrm{od}}=[\nu(\mathbf{r},\mathbf{p})-\nu(\mathbf{r},-\mathbf{p})]/2$.
$E=E^\mathrm{ev}+E^\mathrm{od}$ is the Bogoliubov quasi-particle dispersion relation, with $E^\mathrm{ev}=\sqrt{\left(h_R^{\mathrm{ev}}\right)^2+\Delta_R^2}$ and $E^\mathrm{od}=h_R^\mathrm{od}$.
The quasi-particle distribution function $\nu=\nu^\mathrm{ev}+\nu^\mathrm{od}$ is defined through
\begin{equation}
    \rho^\mathrm{ev}=\frac{1}{2}-\frac{h_R^\mathrm{ev}}{2E^\mathrm{ev}}(1-2\nu^\mathrm{ev}),\quad
    \rho^\mathrm{od}=\nu^\mathrm{od},
\label{rhoev_to_nu}
\end{equation}
where $\nu^\mathrm{ev}$ also relates to the real part of pair coherence $\kappa_R\equiv\mathrm{Re}(\kappa)$,
\begin{equation}
    \kappa_R={\Delta_R(2\nu^\mathrm{ev}-1)}/{2E^\mathrm{ev}}.
\label{kappa_to_nu}
\end{equation}

Note that as we drop the $\mathcal{O}(\hbar^2)$ terms during the Wigner transformation, we neglect the dynamics induced by elastic collisions.
Fortunately, in the BCS limit
$
-1\ll \mathrm{Re}(a_s)\,k_F<0,
$
the characteristic time scale of the elastic dynamics is exponentially long,
$
t_{\rm elastic}\;\sim\;{1}/{\Delta_R(0)}\;\propto\;\exp\!\Big[{\pi}/{2|\mathrm{Re}(a_s)|\,k_F}\Big],
$
where $\Delta_R(0)$ is the initial gap~\cite{yuzbashyan2006dynamical}. By contrast, the dissipative evolution driven by one- and two-body losses unfolds on algebraic time scales,
$
t_{\rm diss}\;\sim\;{1}/{\gamma}
\quad\text{or}\quad
t_{\rm diss}\;\sim\;-\,{1}/{\mathrm{Im}(a_s)}\,,
$
as in Eq.~(\ref{critical_time}).  Hence, for any finite and non-negligible loss rate, the $\mathcal{O}(\hbar)$ dissipative dynamics dominate over the exponentially slow elastic dynamics.  This clear separation of time scales justifies neglecting the $\mathcal{O}(\hbar^2)$ terms and is also the reason that dissipative dynamics are universal in the BCS regime.

Without $h_I$, Eq.~(\ref{nu_IQBE}) reduces to a Vlasov equation of quasi-particle distribution, which has been derived in the literature~\cite{betbeder-matibet1969transporta,urban2006dynamics}.
The addition of one-body and two-body loss ($h_I$) couples the dynamics of quasi-particle distribution $\nu$ to the evolution of $h^\mathrm{ev}$ and $\Delta_R$, or equivalently, the time-even normal particle distribution $\rho^\mathrm{ev}$.
Under the gauge choice $\mathrm{Im}(\Delta_R)=0$ [with the gap equation Eq.~(\ref{gap_eqn}), and the quasi-particle relations Eq.~(\ref{rhoev_to_nu}-\ref{kappa_to_nu})], we find the evolution of $\rho^\mathrm{ev}$ follows,
\begin{equation}
    \int d^3p [\dot{\rho}^\mathrm{ev}-\{h_R^\mathrm{ev},\rho^\mathrm{od}\}_c-\{h_R^\mathrm{od},\rho^\mathrm{ev}\}_c-2h_I\rho^\mathrm{ev}+2\Delta_I\kappa_R]=0.\label{continuity_eqn}
\end{equation}

\begin{figure}[t!]
    \centering
    \includegraphics[width=0.499\textwidth]{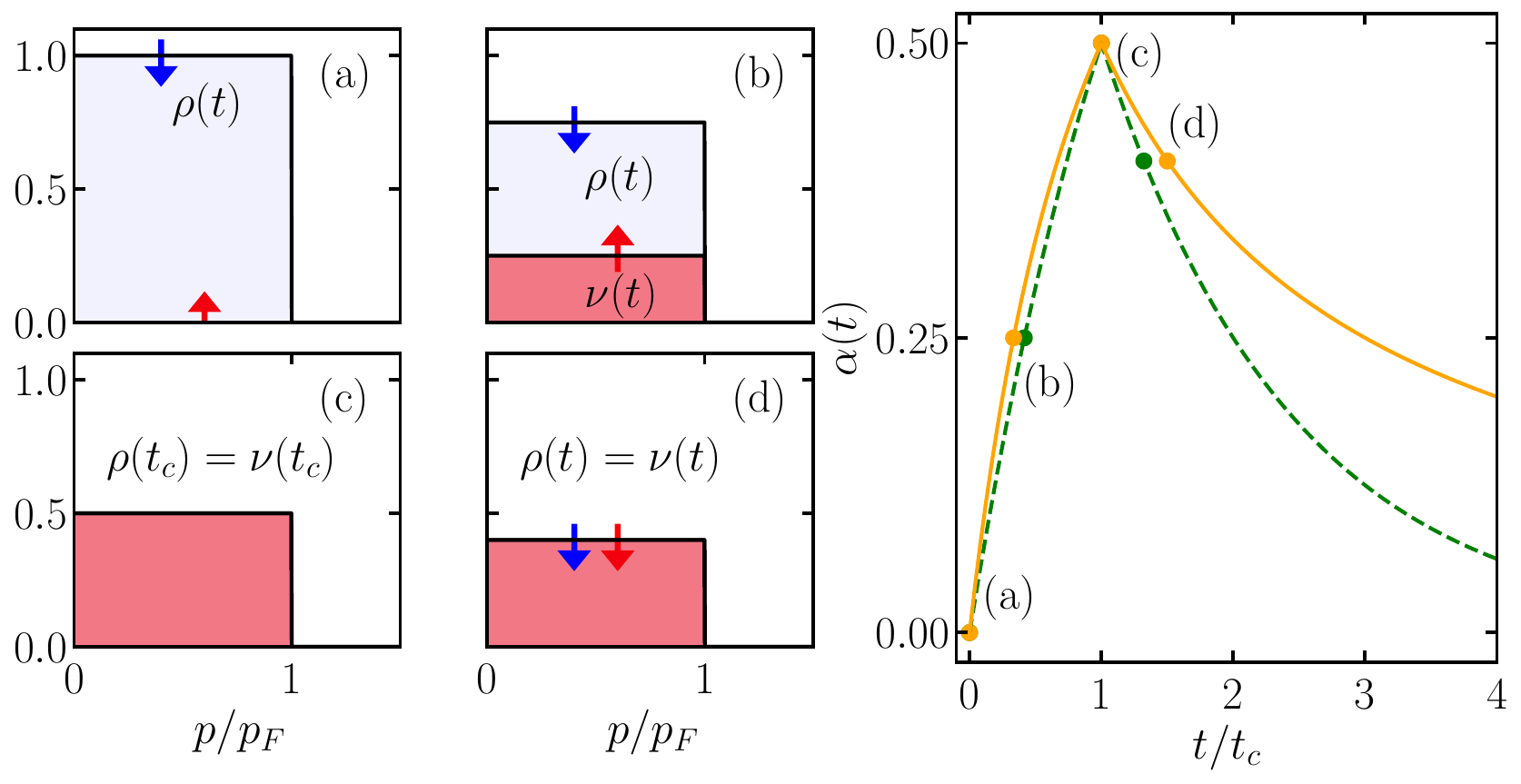}
    \caption{
        Left panel (a-d): schematics of the time evolution of dimensionless quasi-particle ($p_F\nu$) and physical particle ($p_F\rho$) distributions of a homogeneous system in the BCS limit.
Arrows indicate the trend of change.
The critical time $t_c$ is defined in Eq.~(\ref{critical_time}).
Right panel: amplitude of quasi-particle distribution $\alpha$ as a function of time.
Green dashed and orange solid lines are for the system with one- and two-body loss, respectively.}
    \label{fig2}
\end{figure}

\textit{Solution in Homogeneous Systems, BCS limit.---}
Generally, the lossy dynamics described by the above framework can only be solved numerically.
Nevertheless, in the case of homogeneous systems and the BCS limit, we find an analytical solution that helps demonstrate the basic features of such dynamics.
As all quantities are independent of $\mathbf{r}$, Poisson brackets evaluate to zero, and only time-even quantities remain (we omit the superscript for convenience).
We begin with an equilibrated superfluid in the BCS limit, where $T\rightarrow0$ and $\Delta_R\rightarrow0^+$.
Consequently, the equation of motion for Bogoliubov quasiparticle [Eqs.~(\ref{nu_IQBE})] simplifies to
\begin{align}
   \dot{\nu}&=\left(-2\gamma+2g_I\int\frac{d^3p}{(2\pi\hbar)^3}\rho\right)\frac{2\nu^2-2\nu+\rho}{2\nu-1},\label{nu_IQBE_simplified}
\end{align}
with initial conditions $\nu(|\mathbf{p}|<p_F,t=0)p_F\simeq 0$ and $\rho(|\mathbf{p}|<p_F,t=0)p_F\simeq1$, where $p_F=\hbar k_F=\hbar[6\pi^2n(0)]^{1/3}$ is the Fermi momentum of the initial system.
In the BCS limit $\Delta_{R/I}\rightarrow0$, the coherence enhancement of two-body loss is negligible; thus the gauge constraint for the density $\rho$ [Eq.~(\ref{continuity_eqn})] simplifies to
\begin{equation}
    \int d^3p(\dot\rho-2h_I\rho)=0.
\end{equation}

We find that the following solution obeys all required equations:
\begin{equation}
\begin{aligned}
    \nu=\frac{\alpha(t)}{p_F}\theta(1-p/p_F),\quad \rho=\frac{1-\alpha(t)}{p_F}\theta(1-p/p_F).
\end{aligned}
\end{equation}
When the system only has one-body or two-body loss, i.e., $g_I\rightarrow0$ or $\gamma\rightarrow 0$, the quasi particle distribution amplitude $\alpha(t)$ simply reads 
\begin{align}
    &\alpha(t;g_I=0)=\frac{1}{2}-\frac{1}{2}e^{-2\gamma t}|e^{2\gamma t}-2|,\label{nu_with_zero_gI}\\
    &\alpha(t;\gamma=0)=\frac{1}{2}-\frac{|g_I^2k_F^6t^2-9\pi^4|}{2(g_Ik_F^3t+3\pi^2)^2}.\label{nu_with_zero_gamma}
\end{align}

Figure~\ref{fig2} shows the evolution of real ($\rho$) and quasi-particle ($\nu$) distribution as a function of time.
Both Eqs.~(\ref{nu_with_zero_gI}) and (\ref{nu_with_zero_gamma}) lead to similar behavior in $\nu$: it increases monotonically at short times, and suddenly decreases at a critical time $t_c$, then coincides with $\rho$ thereafter.
Below, we explain it using BCS theory.
Initially, the system is in the ground state without any excitation, thus $\nu=0$ [Fig.~\ref{fig2}(a)].
The dissipation of particles below the Fermi surface creates hole excitations, increasing the number of Bogoliubov quasi-particles and thus $\nu$ [Fig.~\ref{fig2}(b)].
In the long-time regime, the values of $\nu$ and $\rho$ coincide, indicating that Bogoliubov quasi-particles become indistinguishable from physical particles [Fig.~\ref{fig2}(d)].

\begin{figure}[t!]
    \centering
    \includegraphics[width=0.499\textwidth]{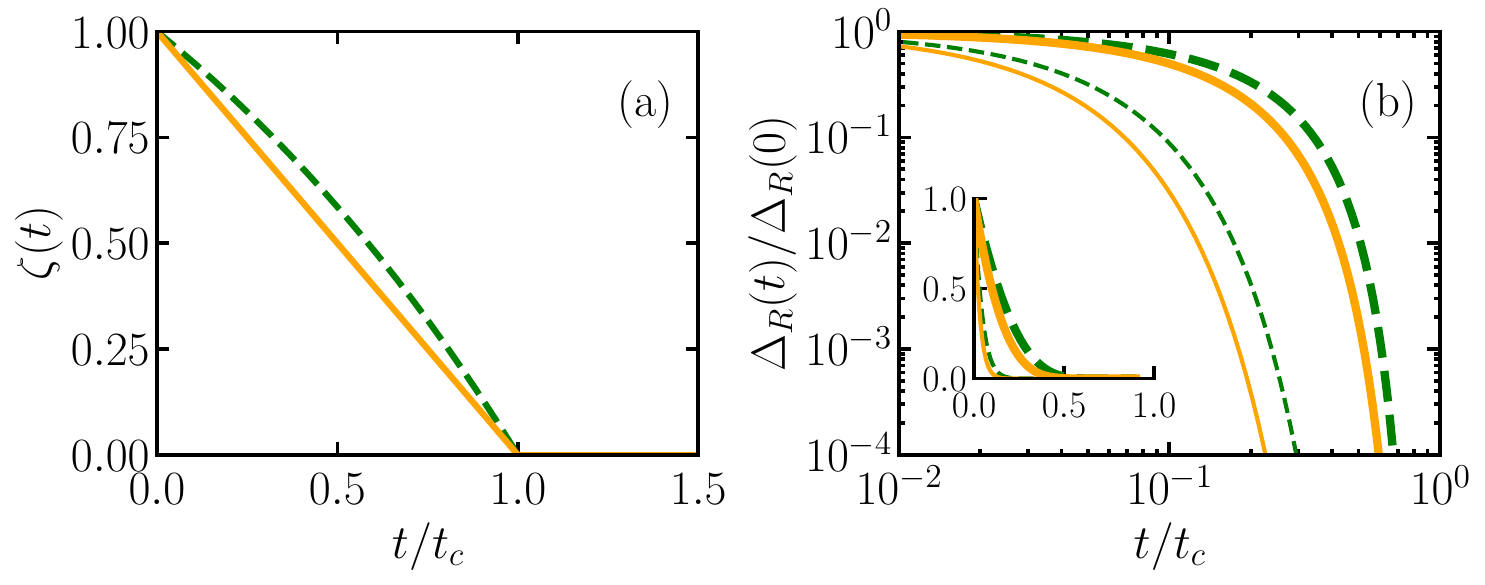}
    \caption{Dynamics of (a) superfluid fraction and (b) gap.
Green dashed and orange solid lines are for systems with one- and two-body loss, respectively.
In (b), thicker and thinner lines are for initial conditions with $\mathrm{Re}(a_s)k_F=-0.5$ and $\mathrm{Re}(a_s)k_F=-0.1$, respectively.
Inset of (b) shows the same results on linear scales.}
    \label{fig3}
\end{figure}

To verify this interpretation, we further calculate the dynamics of superfluid fraction $\zeta$, which can be obtained from the quasiparticle distribution $\nu$ by~\cite{lifshitz2013statistical,fukushima2007superfluid}
\begin{equation} 
\zeta(t)=1+\frac{1}{3m n(t)}\int_0^\infty\frac{d^3p}{(2\pi\hbar)^3} p^2 \frac{\partial \nu}{\partial E^\mathrm{ev}}.
\label{superfluid_fraction}
\end{equation}
We find the system completely losses its superfluidity at $\alpha(t_c)=1/2$, where the critical time $t_c$ is
\begin{equation}
    t_c(g_I=0)=\ln(2)/2\gamma,\quad t_c(\gamma=0)=-3\pi^2/g_Ik_F^3.
\label{critical_time}
\end{equation} The full analytical form of the superfluid fraction reads (See Supplemental Material)
\begin{equation}
    \zeta(t) = \begin{cases}
        {[1-2\alpha(t)]}/{[1-\alpha(t)]} & t<t_c,\\
        0 & t\ge t_c.
\end{cases}
    \label{superfluid_fraction_dynamics}
\end{equation}
It is clear that the first-order derivative of superfluid fraction with respect to $t$ is also discontinuous at $t_c$ (Fig.~\ref{fig3}), indicating a dynamical phase transition with a critical time $t_c$.
Note that, first, because the system's evolution is always under non-equilibrium states where temperature is ill-defined, the time cannot be simply reinterpreted as an imaginary temperature.
Second, the critical time $t_c$ depends solely on $\gamma$ and $\mathrm{Im}(a_s)$, independent of $\mathrm{Re}(a_s)$.
This is because $\nu(k<k_F,t)$ is not influenced by $\mathrm{Re}(a_s)$ [see Eq.~(\ref{nu_IQBE_simplified})].

Likewise, from the gap equation Eq.~(\ref{gap_eqn}), we find that the gap $\Delta_R(t)$ also reaches 0 at $t_c$:
\begin{equation}
    \Delta_R(t)=\begin{cases}
    \dfrac{8 E_F}{e^2}\exp\left\{\dfrac{\pi}{2k_F \mathrm{Re}(a_s)[1-2\alpha(t)]}\right\}&t<t_c\\
    0&t\geq t_c,
    \end{cases}
    \label{order_parameter_dynamics}
\end{equation}
where $E_F=p_F^2/2m$ is the Fermi energy.
While $\mathrm{Re}(a_s)$ does not affect $t_c$, it significantly impacts the decay rate.
As shown in Fig.~(\ref{fig3})(b), a larger absolute value of elastic scattering length $|\mathrm{Re}(a_s)|$ slows the decay of $\Delta_R$.
Notably, $\Delta_R$ decays even faster than exponential form, in contrast to the exponential decay in Hermitian dynamical vanishing of the order parameter~\cite{yuzbashyan2006dynamical}. While it appears that the decay of the gap is a smooth function, it is non-analytic at $t_c$, which resembles the case for the thermal Berezinskii-Kosterlitz-Thouless transition. From Eq.~(\ref{kappa_to_nu}), it is clear that the anomalous density matrix $\kappa_R$ also vanishes at $t_c$, and the generalized density matrix reduces to the normal density matrix $\rho$, of which the dynamics are further described by the equation of motion of the normal fluids [Eq.~(\ref{TDHF})].

We emphasize that our dissipative dynamical phase transition is universal for a BCS superfluid subject to one-body or two-body dissipation. That is, starting from a BCS ground state with any elastic scattering length, the system will always undergo a dynamical phase transition at a critical time $t_c$.
This is in contrast to the typical dynamical phase transition for closed systems, where the initial state has to be carefully engineered~\cite{heyl2013dynamical,heyl2018dynamical}.

\textit{Conclusion.---}
To summarize, we generalized the time-dependent Hartree-Fock-Bogoliubov equation to describe lossy fermionic superfluids by introducing a variational method based on the least action principle, which extends beyond the limitations of steady-state analyses.
Under the semi-classical approximation, we derived equations governing the evolution of Bogoliubov quasi-particle distributions.
For homogeneous systems in the BCS limit, we obtained solutions that reveal a universal dynamical phase transition: lossy fermionic superfluids exit the superfluid phase at a critical time with a faster-than-exponential and non-analytic decay of the order parameter.
This result highlights the profound impact of dissipation on phase transitions.
Our work opens up new avenues for modeling open quantum systems beyond steady-state scenarios, which is versatile and can be applied to a wide range of systems.

\begin{acknowledgments}
We thank Chenwei Lv, Ren Zhang, Zhe-Yu Shi, and Kaiyuen Lee for the helpful discussion.
We would like to acknowledge financial support from the National Natural Science Foundation of China under Grant No. 124B2074 and No. 12204395, 
Hong Kong RGC Early Career Scheme (Grant No. 24308323) and Collaborative Research Fund (Grant No. C4050-23GF), 
the Space Application System of China Manned Space Program, 
Guangdong Provincial Quantum Science Strategic Initiative GDZX2404004, 
and CUHK Direct Grant No. 4053731.
\end{acknowledgments}

\appendix
\section{End Matter}
\subsection{Master Equation and Adjoint Equation}
The action Eq.~(\ref{S}) is defined in a time interval from the initial time $t_i$ to the final time $t_f$.
Importantly, the boundary condition for $\widehat{D}$ is provided at the initial time, while for $\widehat{A}$, it is provided at the final time, i.e.
$\widehat{D}(t_i)=\widehat{D}_i$ and $\widehat{A}(t_f)=\widehat{A}_f$.
By the least action principle, $\delta S/\delta \widehat{A}=0$ yields

\begin{equation}
\begin{aligned}
    &\int_{t_i}^{t_f}dt \mathrm{Tr} \delta \widehat{A} \bigg[i\hbar\frac{d\widehat{D}}{dt}-[\widehat{H},\widehat{D}]\\
    &-i\hbar\sum_n\left(\frac{1}{2}\{\widehat{L}^\dagger_n\widehat{L}_n,\widehat{D}\}-\widehat{L}_n\widehat{D}\widehat{L}^\dagger_n\right)\bigg]=0.
\label{var_lindblad_eqn}
\end{aligned}
\end{equation}

If $\delta \widehat{A}$ takes arbitrary variation, Eq.~(\ref{var_lindblad_eqn}) produces the equation of motion of $\widehat{D}(t-t_i)$, which is the Lindblad master equation~\cite{lindblad1976generatorsa,daley2014quantum}.
$\delta S/\delta \widehat{D}=0$ yields

\begin{equation}
\begin{aligned}
    &\int_{t_i}^{t_f}dt \mathrm{Tr} \delta \widehat{D} \bigg[i\hbar\frac{d\widehat{A}}{dt}-[\widehat{H},\widehat{A}]\\
    &-i\hbar\sum_n\left(\frac{1}{2}\{\widehat{L}^\dagger_n\widehat{L}_n,\widehat{A}\}-\widehat{L}_n^\dagger\widehat{A}\widehat{L}_n\right)\bigg]=0.
\end{aligned}
\label{adjoint_eqn}
\end{equation}
If $\delta \widehat{D}$ takes arbitrary variation, Eq.~(\ref{var_lindblad_eqn}) produces an auxiliary equation of motion of $\widehat{A}(t_f-t)$, which is usually referred to as adjoint equation~\cite{breuer2002theory,gardiner2004quantum,roberts2021hidden}.
It should be emphasized that the physical interpretation of adjoint equation can only be obtained by extracting expectation value $\langle\widehat{A}(t)\rangle=\mathrm{Tr}[\widehat{D}_i\widehat{A}(t)]$ rather than directly interpreting $\widehat{A}(t)$, which is only of an auxiliary usage, since the latter may break canonical (anti)-commutation relations~\cite{fazio2024manybody}.

The key advantage of this reformulation of both master equation and adjoint equation lies in its flexibility in solving the general question of obtaining the final time expectation of $\widehat{A}_f$ using various approximation methods: One can choose arbitrary ansatz of $\widehat{D}(t)\approx \widehat{D}_\mathrm{a}(\mathbf{c}^D(t))$ with $\widehat{D}_i=\widehat{D}_\mathrm{a}(\mathbf{c}^D(t_i))$, and $\widehat{A}(t)\approx \widehat{A}_\mathrm{a}(\mathbf{c}^A(t))$ with $\widehat{A}_f=\widehat{A}_\mathrm{a}(\mathbf{c}^A(t_f))$ to insert into Eq.~(\ref{S}), where $\mathbf{c}^D$ and $\mathbf{c}^A$ are parameters in the ansatz.
Then, applying the least action principle directly yields effective master and adjoint equations to be
\begin{equation}
    \frac{d S[\widehat{D}_\mathrm{a},\widehat{A}_\mathrm{a}]}{d \mathbf{c}^A}=0,~~~\frac{d S[\widehat{D}_\mathrm{a},\widehat{A}_\mathrm{a}]}{d \mathbf{c}^D}=0.\label{eff_eqns}
\end{equation}
Generally, there will be couplings between two equations in Eqs.~(\ref{eff_eqns}).
By solving them simultaneously, one obtains $\widehat{D}_\mathrm{a}(\mathbf{c}^D(t_f))$ then the expectation of $\widehat{A}_f$ under certain approximation.
In some cases, by properly choosing the ansatz, Eqs.~(\ref{eff_eqns}) may decouple, leading to much simplification.
The Hartree-Fock-Bogoliubov approximation discussed in this work is an example.

\subsection{General Quadratic Form of Ansatz $\widehat{A}_a$}
Explicitly, $\widehat{A}_a$ is set to be
\begin{equation}
\begin{aligned}
    &\widehat{A}_\mathrm{a}(t)=\sum_{\mathbf{k},\sigma,\mathbf{k}',\sigma'}\big[a_{\mathbf{k},\sigma, \mathbf{k}',\sigma'}(t)\left(c^\dagger_{\mathbf{k},\sigma}c_{\mathbf{k}',\sigma'}-c_{\mathbf{k}',\sigma'}c^\dagger_{\mathbf{k},\sigma}\right)\\
    &+b_{\mathbf{k},\sigma, \mathbf{k}',\sigma'}(t)c^\dagger_{\mathbf{k},\sigma}c^\dagger_{\mathbf{k}',\sigma'}+b_{\mathbf{k},\sigma, \mathbf{k}',\sigma'}^*(t)c_{\mathbf{k}',\sigma'}c_{\mathbf{k},\sigma}\big],
\end{aligned}
\label{Aa}
\end{equation}
where we emphasize that we assign all temporal dependence on ansatz parameters $a$ and $b$. The reason to choose such form is that we want to measure two types of observables, field coherence $c^\dagger_{\mathbf{k},\sigma}c_{\mathbf{k}',\sigma'}$ and pairing coherence $c_{\mathbf{k}',\sigma'}c_{\mathbf{k},\sigma}$.
Therefore, we need to ensure $\widehat{A}_\mathrm{a}$ has a general form, which can always be tuned to fit all possible desired $\widehat{A}_f$.

\subsection{Effective Adjoint Equation}
In the main text, we show that the effective equation of motion obtained by $\delta S[\mathcal{R},\mathcal{A}]/\delta \mathcal{A}=0$ is the generalized time-dependent Hartree-Fock-Bogoliubov equation Eq.~(\ref{TDHFB}).
For completeness, we consider another variation relation $\delta S[\mathcal{R},\mathcal{A}]/\delta \mathcal{R}=0$, which yields
\begin{equation}
\begin{aligned}
    i\hbar \dot{\mathcal{A}}=&[\mathcal{H}_R,\mathcal{A}]\left(\frac{\delta \mathcal{H}_R}{\delta R}+\mathbb{I}\right)\\
    &+i\hbar\{\mathcal{H}_I,\mathcal{A}\}\left(\frac{\delta \mathcal{H}_I}{\delta R}+\mathbb{I}\right)+i\hbar \mathcal{A}\frac{\delta \mathcal{J}}{\delta R}.
\label{EOM_A}
\end{aligned}
\end{equation}
where we denote $\mathbb{I}$ to be the identity matrix in the single-particle space.
It is noted that unlike Eq.~(\ref{TDHFB}), the effective equation of motion of $\mathcal{A}$ Eq.~(\ref{EOM_A}) does not close by itself, because $\mathcal{H}_R[R], \mathcal{H}_I[R]$ and $\mathcal{J}[R]$ all depend on $\mathcal{R}(t)$.

\clearpage
\widetext
\begin{center}
\textbf{\large Supplemental Material: Dynamical Phase Transition of Dissipative Fermionic Superfluids}
\end{center}
%%%%%%%%%% Merge with supplemental materials %%%%%%%%%%
%%%%%%%%%% Prefix a "S" to all equations, figures, tables and reset the counter %%%%%%%%%%
\setcounter{equation}{0}
\setcounter{figure}{0}
\setcounter{table}{0}
\setcounter{page}{8}
\makeatletter
\renewcommand{\theequation}{S\arabic{equation}}
\renewcommand{\thefigure}{S\arabic{figure}}
\renewcommand{\bibnumfmt}[1]{[S#1]}
\renewcommand{\thesection}{S\@arabic\c@section}
% \renewcommand{\citenumfont}[1]{S#1}
%%%%%%%%%% Prefix a "S" to all equations, figures, tables and reset the counter %%%%%%%%%%

%\tableofcontents

\section{Hartree-Fock-Bogoliubov Approximation on Auxiliary Observable Action}

In this section, we provide details on obtaining Eq.~(\ref{Seff}) under Hartree-Fock-Bogoliubov approximation from Eq.~(\ref{S}). For convenience, let's consider the constituents of $\widehat{A}_\mathrm{a}$: 
\begin{equation}
\begin{aligned}
    &\widehat{A}_{11}=a_{\mathbf{k},\mathbf{k}'}c^\dagger_{\mathbf{k},\uparrow}c_{\mathbf{k}',\uparrow},\quad \widehat{A}_{12}=b_{\mathbf{k},\mathbf{k}'}^*c_{\mathbf{k}',\downarrow}c_{\mathbf{k},\uparrow},\\
    &\widehat{A}_{21}=\widehat{A}^{\dagger}_{12},\quad \widehat{A}_{22}=\widehat{A}_{11}-a_{\mathbf{k},\mathbf{k}'}\delta_{\mathbf{k},\mathbf{k}'}.
\end{aligned}
\label{A_constituents}
\end{equation}
As mentioned in the main text, because we are dealing with the spin-balanced system, we only consider the field and pairing coherence of one species. From Eq.~(\ref{A_constituents}), we recover $\widehat{A}$ by:
\begin{equation}
    \widehat{A}_\mathrm{a}=\sum_{\mathbf{k},\mathbf{k}'}(\widehat{A}_{11}+\widehat{A}_{12}+\widehat{A}_{21}+\widehat{A}_{22}).
\end{equation}

Let us consider the first term in Eq.~(\ref{S}), based on Eq.~(\ref{A_constituents}),
\begin{equation}
\begin{aligned}
    \hbar\operatorname{Tr}\widehat{D}_\mathrm{a}(t_f)\widehat{A}_\mathrm{a}(t_f)&=\hbar\sum_{\mathbf{k},\mathbf{k}'}(\langle\widehat{A}_{11}\rangle+\langle\widehat{A}_{12}\rangle+\langle\widehat{A}_{21}\rangle+\langle\widehat{A}_{22}\rangle)\\
    &=\hbar\sum_{\mathbf{k},\mathbf{k}'}( 2\langle\widehat{A}_{11}\rangle-a_{\mathbf{k},\mathbf{k}'}\delta_{\mathbf{k},\mathbf{k}'}+2\mathrm{Re}\langle\widehat{A}_{12}\rangle),
\end{aligned}
\end{equation}
All temporal dependence above is set to be on $t_f$. According to Wick's theorem, 
\begin{align}
    &\langle\widehat{A}_{11}\rangle=a_{\mathbf{k},\mathbf{k}'}\langle c^\dagger_{\mathbf{k},\uparrow}c_{\mathbf{k}',\uparrow}\rangle=a_{\mathbf{k},\mathbf{k}'}\rho_{\mathbf{k}',\mathbf{k}},\\
    &\langle\widehat{A}_{12}\rangle=b_{\mathbf{k},\mathbf{k}'}^*\langle c_{\mathbf{k}',\downarrow}c_{\mathbf{k},\uparrow}\rangle=b_{\mathbf{k},\mathbf{k}'}^*\kappa_{\mathbf{k},\mathbf{k}'}.
\end{align} 
Consequently,
\begin{equation}
    \hbar\operatorname{Tr}\widehat{D}_\mathrm{a}(t_f)\widehat{A}_\mathrm{a}(t_f)=\hbar\sum_{\mathbf{k},\mathbf{k}'}(2a_{\mathbf{k},\mathbf{k}'}\rho_{\mathbf{k}',\mathbf{k}}-a_{\mathbf{k},\mathbf{k}'}\delta_{\mathbf{k},\mathbf{k}'}+b_{\mathbf{k},\mathbf{k}'}^*\kappa_{\mathbf{k},\mathbf{k}'}+b_{\mathbf{k},\mathbf{k}'}\kappa_{\mathbf{k},\mathbf{k}'}^*).\label{first_term_1}
\end{equation}
Because we require $\widehat{A}$ and $\widehat{W}$ to be Hermitian, it is straightforward to show $a$ and $\rho$ are Hermitian, $b$ and $\kappa$ are skew-symmetric:
\begin{equation}
\begin{aligned}
    &a_{\mathbf{k},\mathbf{k}'}=a^\dagger_{\mathbf{k},\mathbf{k}'},\quad \rho_{\mathbf{k},\mathbf{k}'}=\rho^\dagger_{\mathbf{k},\mathbf{k}'}\\
    &b_{\mathbf{k},\mathbf{k}'}=-b_{\mathbf{k}',\mathbf{k}},\quad \kappa_{\mathbf{k},\mathbf{k}'}=-\kappa_{\mathbf{k}',\mathbf{k}}.
\end{aligned}
\end{equation}
Thus, we can rearrange Eq.~(\ref{first_term_1}) into
\begin{equation}
\begin{aligned}
    \hbar\operatorname{Tr}\widehat{D}_\mathrm{a}(t_f)\widehat{A}_\mathrm{a}(t_f)=&\sum_{\mathbf{k}}\delta_{\mathbf{k},\mathbf{k}'}\left(\sum_{\mathbf{k}''}\rho_{\mathbf{k},\mathbf{k}''}a_{\mathbf{k}'',\mathbf{k}'}+\kappa_{\mathbf{k},\mathbf{k}''}b^\dagger_{\mathbf{k}'',\mathbf{k}'}+\kappa^\dagger_{\mathbf{k},\mathbf{k}''}b_{\mathbf{k}'',\mathbf{k}'}+\rho^*_{\mathbf{k},\mathbf{k}''}a^*_{\mathbf{k}'',\mathbf{k}'}-\delta_{\mathbf{k},\mathbf{k}''}a^*_{\mathbf{k}'',\mathbf{k}'}\right)\\
    &=\tr\hbar\mathcal{R}(t_f)\mathcal{A}(t_f).
\end{aligned}
\end{equation}
Similarly, we can prove the second term in Eq.~(\ref{S}) in the main text has the relation
\begin{equation}
    \Tr\hbar\widehat{A}_\mathrm{a}(t)\frac{d \widehat{D}_\mathrm{a}(t)}{dt}=\tr\hbar\mathcal{A}(t)\frac{d\mathcal{R}(t)}{dt}.
\end{equation}
Then, let us consider the third term. According to Eq.~(\ref{A_constituents}),
\begin{equation}
\begin{aligned}
    \Tr i\widehat{A}_\mathrm{a}[\widehat{H},\widehat{D}_\mathrm{a}]= i\langle[\widehat{A}_\mathrm{a},\widehat{H}]\rangle=i\sum_{\mathbf{k},\mathbf{k}'}(2\langle[\widehat{A}_{11},\widehat{H}]\rangle+2\mathrm{Re}\langle[\widehat{A}_{12},\widehat{H}]\rangle). 
\end{aligned}
\label{AHD1}
\end{equation}
All temporal dependence above is set to be on $t_i<t<t_f$. We apply Wick's theorem for $\langle[\widehat{A}_{11},\widehat{H}]\rangle$ and $\langle[\widehat{A}_{12},\widehat{H}]\rangle$ separately, which yields
\begin{equation}
\begin{aligned}
    \langle[\widehat{A}_{11},\widehat{H}]\rangle=&\sum_\mathbf{q} a_{\mathbf{k},\mathbf{k}'} U(|\mathbf{k}'-\mathbf{q}|)(\rho_{\mathbf{q},\mathbf{k}}-\rho_{\mathbf{k},\mathbf{q}})\\
    &+\frac{a_{\mathbf{k},\mathbf{k}'}g_R}{V}\sum_{\mathbf{P},\mathbf{q},\mathbf{q}'}\delta_{\mathbf{k}',\frac{\mathbf{P}}{2}-\mathbf{q}}\rho_{\frac{\mathbf{P}}{2}-\mathbf{q}',\mathbf{k}}\rho_{\frac{\mathbf{P}}{2}+\mathbf{q}',\frac{\mathbf{P}}{2}+\mathbf{q}}+\frac{a_{\mathbf{k},\mathbf{k}'}g_R}{V}\sum_{\mathbf{P},\mathbf{q},\mathbf{q}'}\delta_{\mathbf{k}',\frac{\mathbf{P}}{2}-\mathbf{q}}\kappa^*_{\mathbf{k},\frac{\mathbf{P}}{2}+\mathbf{q}}\kappa_{\frac{\mathbf{P}}{2}-\mathbf{q}',\frac{\mathbf{P}}{2}+\mathbf{q}'}\\
    &-\frac{a_{\mathbf{k},\mathbf{k}'}g_R}{V}\sum_{\mathbf{P},\mathbf{q},\mathbf{q}'}\delta_{\mathbf{k},\frac{\mathbf{P}}{2}-\mathbf{q}'}\rho_{\mathbf{k}',\frac{\mathbf{P}}{2}-\mathbf{q}}\rho_{\frac{\mathbf{P}}{2}+\mathbf{q}',\frac{\mathbf{P}}{2}+\mathbf{q}}-\frac{a_{\mathbf{k},\mathbf{k}'}g_R}{V}\sum_{\mathbf{P},\mathbf{q},\mathbf{q}'}\delta_{\mathbf{k},\frac{\mathbf{P}}{2}-\mathbf{q}'}\kappa^*_{\frac{\mathbf{P}}{2}-\mathbf{q},\frac{\mathbf{P}}{2}+\mathbf{q}}\kappa_{\mathbf{k}',\frac{\mathbf{P}}{2}+\mathbf{q}'},
\end{aligned}
\label{A11H}
\end{equation}
and
\begin{equation}
\begin{aligned}
    \langle[\widehat{A}_{12},\widehat{H}]\rangle=&(\epsilon_\mathbf{k}+\epsilon_{\mathbf{k}'}-2\mu)b_{\mathbf{k},\mathbf{k}'}^*\kappa_{\mathbf{k},\mathbf{k}'}+\sum_\mathbf{q}b_{\mathbf{k},\mathbf{k}'}^*\left[U(|\mathbf{k}-\mathbf{q}|)\kappa_{\mathbf{q},\mathbf{k}'}+U(|\mathbf{k}'-\mathbf{q}|)\kappa_{\mathbf{k},\mathbf{q}}\right]\\
    &+\frac{b_{\mathbf{k},\mathbf{k}'}^*g_R}{V}\sum_{\mathbf{P},\mathbf{q},\mathbf{q}'}\delta_{\mathbf{k},\frac{\mathbf{P}}{2}-\mathbf{q}}\delta_{\mathbf{k}',\frac{\mathbf{P}}{2}+\mathbf{q}}\kappa_{\frac{\mathbf{P}}{2}-\mathbf{q}',\frac{\mathbf{P}}{2}+\mathbf{q}'}\\
    &-\frac{b_{\mathbf{k},\mathbf{k}'}^*g_R}{V}\sum_{\mathbf{P},\mathbf{q},\mathbf{q}'}\delta_{\mathbf{k}',\frac{\mathbf{P}}{2}+\mathbf{q}}\rho_{\frac{\mathbf{P}}{2}-\mathbf{q}',\frac{\mathbf{P}}{2}-\mathbf{q}}\kappa_{\frac{\mathbf{P}}{2}+\mathbf{q}',\mathbf{k}}-\frac{b_{\mathbf{k},\mathbf{k}'}^*g_R}{V}\sum_{\mathbf{P},\mathbf{q},\mathbf{q}'}\delta_{\mathbf{k}',\frac{\mathbf{P}}{2}+\mathbf{q}}\rho_{\mathbf{k},\frac{\mathbf{P}}{2}-\mathbf{q}}\kappa_{\frac{\mathbf{P}}{2}-\mathbf{q}',\frac{\mathbf{P}}{2}+\mathbf{q}'}\\
    &-\frac{b_{\mathbf{k},\mathbf{k}'}^*g_R}{V}\sum_{\mathbf{P},\mathbf{q},\mathbf{q}'}\delta_{\mathbf{k},\frac{\mathbf{P}}{2}-\mathbf{q}}\rho_{\frac{\mathbf{P}}{2}+\mathbf{q}',\frac{\mathbf{P}}{2}+\mathbf{q}}\kappa_{\frac{\mathbf{P}}{2}-\mathbf{q}',\mathbf{k}'}-\frac{b_{\mathbf{k},\mathbf{k}'}^*g_R}{V}\sum_{\mathbf{P},\mathbf{q},\mathbf{q}'}\delta_{\mathbf{k},\frac{\mathbf{P}}{2}-\mathbf{q}}\rho_{\mathbf{k}',\frac{\mathbf{P}}{2}+\mathbf{q}}\kappa_{\frac{\mathbf{P}}{2}-\mathbf{q}',\frac{\mathbf{P}}{2}+\mathbf{q}'}.
\end{aligned}
\label{A12H}
\end{equation}
In the above calculation, we do not distinguish the contribution from different spin species based on the relation $\langle c^\dagger_{\mathbf{k}',\uparrow}c_{\mathbf{k},\uparrow}\rangle=\langle c^\dagger_{\mathbf{k}',\downarrow}c_{\mathbf{k},\downarrow}\rangle = \rho_{\mathbf{k},\mathbf{k}'}$ and $\langle c_{\mathbf{k}',\downarrow}c_{\mathbf{k},\uparrow}\rangle=\langle c_{\mathbf{k}',\uparrow}c_{\mathbf{k},\downarrow}\rangle=\kappa_{\mathbf{k},\mathbf{k}'}$. 

To further proceed, we compute the Hartree-Fock Hamiltonians $h_{R}$ and the pairing fields $\Delta_{R}$. In these calculations, we need to emphasize the contribution of different species because, when extracting those quantities for only one species, the functional variation should be against that species only. To conveniently do that, we further define $\rho_{\uparrow,\mathbf{k},\mathbf{k}'}\equiv\rho_{\mathbf{k},\uparrow,\mathbf{k}',\uparrow}=\langle c^\dagger_{\mathbf{k}',\uparrow}c_{\mathbf{k},\uparrow}\rangle$, $\rho_{\downarrow,\mathbf{k},\mathbf{k}'}\equiv\rho_{\mathbf{k},\downarrow,\mathbf{k}',\downarrow}=\langle c^\dagger_{\mathbf{k}',\downarrow}c_{\mathbf{k},\downarrow}\rangle$ and $\kappa_{\uparrow,\mathbf{k},\mathbf{k}'}\equiv \kappa_{\mathbf{k},\uparrow,\mathbf{k}',\downarrow}=\langle c_{\mathbf{k}',\downarrow}c_{\mathbf{k},\uparrow}\rangle$, $\kappa_{\downarrow,\mathbf{k},\mathbf{k}'}\equiv \kappa_{\mathbf{k},\downarrow,\mathbf{k}',\uparrow}=\langle c_{\mathbf{k}',\uparrow}c_{\mathbf{k},\downarrow}\rangle$.
Applying Wick's theorem to the energy functional $\operatorname{Tr}(\widehat{D}\widehat{H}_{R})\equiv\langle\widehat{H}_{R}\rangle$, we obtain
\begin{equation}
\begin{aligned}
    \langle H_R\rangle&=\sum_\mathbf{q}(\epsilon_\mathbf{q}-\mu)(\rho_{\uparrow,\mathbf{q},\mathbf{q}}+\rho_{\downarrow,\mathbf{q},\mathbf{q}})+\sum_{\mathbf{q},\mathbf{q}'}U(|\mathbf{q}-\mathbf{q}'|)(\rho_{\uparrow,\mathbf{q}',\mathbf{q}}+\rho_{\downarrow,\mathbf{q}',\mathbf{q}})\\
    &+\frac{g_R}{V}\sum_{\mathbf{P},\mathbf{q},\mathbf{q}'}\rho_{\uparrow,\frac{\mathbf{P}}{2}-\mathbf{q},\frac{\mathbf{P}}{2}-\mathbf{q}'}\rho_{\downarrow,\frac{\mathbf{P}}{2}+\mathbf{q},\frac{\mathbf{P}}{2}+\mathbf{q}'}+\frac{g_R}{V}\sum_{\mathbf{P},\mathbf{q},\mathbf{q}'}\kappa^*_{\uparrow,\frac{\mathbf{P}}{2}-\mathbf{q}',\frac{\mathbf{P}}{2}+\mathbf{q}'}\kappa_{\uparrow,\frac{\mathbf{P}}{2}-\mathbf{q},\frac{\mathbf{P}}{2}+\mathbf{q}}.
\end{aligned}
\end{equation}
Based on Eq.~(\ref{variation}) in the main text,
\begin{align}
    &(h_R)_{\mathbf{k},\mathbf{k}'}=(\epsilon_\mathbf{k}-\mu)\delta_{\mathbf{k},\mathbf{k}'}+U(|\mathbf{k}-\mathbf{k}'|)+\frac{g_R}{V}\sum_{\mathbf{P},\mathbf{q},\mathbf{q}'}\delta_{\frac{\mathbf{P}}{2}-\mathbf{q}',\mathbf{k}'}\delta_{\frac{\mathbf{P}}{2}-\mathbf{q},\mathbf{k}}\rho_{\frac{\mathbf{P}}{2}+\mathbf{q}',\frac{\mathbf{P}}{2}+\mathbf{q}},\label{hR}\\
    &(\Delta_R)_{\mathbf{k},\mathbf{k}'}=\frac{g_R}{V}\sum_{\mathbf{P},\mathbf{q},\mathbf{q}'}\delta_{\frac{\mathbf{P}}{2}-\mathbf{q},\mathbf{k}}\delta_{\frac{\mathbf{P}}{2}+\mathbf{q},\mathbf{k}'}\kappa_{\frac{\mathbf{P}}{2}-\mathbf{q}',\frac{\mathbf{P}}{2}+\mathbf{q}'},\label{DeltaR}
\end{align}
After taking the partial derivative, we again drop the spin dependence for convenience. From Eqs.~(\ref{AHD1}), (\ref{A11H}), (\ref{A12H}), (\ref{hR}) and (\ref{DeltaR}), we can arrange $\Tr i \widehat{A}_\mathrm{a}[\widehat{H},\widehat{D}_\mathrm{a}]$ into
\begin{equation}
\begin{aligned}
    i\Tr\widehat{A}_\mathrm{a}[\widehat{H},\widehat{D}_\mathrm{a}]=&i\sum_{\mathbf{k}}\delta_{\mathbf{k},\mathbf{k}'}\sum_{\mathbf{q},\mathbf{q}'}a_{\mathbf{k},\mathbf{q}}(h_R)_{\mathbf{q},\mathbf{q}'}\rho_{\mathbf{q}',\mathbf{k}'}+a_{\mathbf{k},\mathbf{q}}(\Delta_R)_{\mathbf{q},\mathbf{q}'}\kappa^\dagger_{\mathbf{q}',\mathbf{k}'}-a_{\mathbf{k},\mathbf{q}}\kappa_{\mathbf{q},\mathbf{q}'}(\Delta_R^\dagger)_{\mathbf{q}',\mathbf{k}'}-a_{\mathbf{k},\mathbf{q}}\rho_{\mathbf{q},\mathbf{q}'}(h_R)_{\mathbf{q}',\mathbf{k}'}\\
    &+b_{\mathbf{k},\mathbf{q}}(\Delta_R^\dagger)_{\mathbf{q},\mathbf{q}'}\rho_{\mathbf{q}',\mathbf{k}'}-b_{\mathbf{k},\mathbf{q}}\kappa^\dagger_{\mathbf{q},\mathbf{q}'}(h_R)_{\mathbf{q}',\mathbf{k}'}-b_{\mathbf{k},\mathbf{q}}(h_R^*)_{\mathbf{q},\mathbf{q}'}\kappa^\dagger_{\mathbf{q}',\mathbf{k}'}-b_{\mathbf{k},\mathbf{q}}(\delta_{\mathbf{q},\mathbf{q}'}-\rho^*_{\mathbf{q},\mathbf{q}'})(\Delta_R^\dagger)_{\mathbf{q}',\mathbf{k}'}\\
    &+b^\dagger_{\mathbf{k},\mathbf{q}}(h_R)_{\mathbf{q},\mathbf{q}'}\kappa_{\mathbf{q}',\mathbf{k}'}+b^\dagger_{\mathbf{k},\mathbf{q}}(\Delta_R)_{\mathbf{q},\mathbf{q}'}(\delta_{\mathbf{q}',\mathbf{k}'}-\rho^*_{\mathbf{q}',\mathbf{k}'})+b^\dagger_{\mathbf{k},\mathbf{q}}\kappa_{\mathbf{q},\mathbf{q}'}(h_R^*)_{\mathbf{q}',\mathbf{k}'}-b^\dagger_{\mathbf{k},\mathbf{q}}\rho_{\mathbf{q},\mathbf{q}'}(\Delta_R)_{\mathbf{q}',\mathbf{k}'}\\
    &-a^*_{\mathbf{k},\mathbf{q}}(\Delta_R^\dagger)_{\mathbf{q},\mathbf{q}'}\kappa_{\mathbf{q}',\mathbf{k}'}+a^*_{\mathbf{k},\mathbf{q}}\kappa^\dagger_{\mathbf{q},\mathbf{q}'}(\Delta_R)_{\mathbf{q}',\mathbf{k}'}+a^*_{\mathbf{k},\mathbf{q}}(h_R^*)_{\mathbf{q},\mathbf{q}'}(\delta_{\mathbf{q}',\mathbf{k}'}-\rho^*_{\mathbf{q}',\mathbf{k}'})-a^*_{\mathbf{k},\mathbf{q}}(\delta_{\mathbf{q},\mathbf{q}'}-\rho^*_{\mathbf{q},\mathbf{q}'})(h_R^*)_{\mathbf{q}',\mathbf{k}'}\\
    =&i\tr\mathcal{A}[\mathcal{H}_R,\mathcal{R}].
\end{aligned}
\end{equation}

Lastly, we are left with the last two terms with jump operators. It is convenient to consider them together. Again, according to Eq.~(\ref{A_constituents}),
\begin{equation}
\begin{aligned}
    &-\hbar\Tr \widehat{A}_\mathrm{a}\sum_n\left(\frac{1}{2}\{\widehat{L}^\dagger_n\widehat{L}_n,\widehat{D}_\mathrm{a}\}-\widehat{L}_n\widehat{D}_\mathrm{a}\widehat{L}^\dagger_n\right)=-\hbar \sum_n \left(\langle\{\widehat{A}_\mathrm{a},\frac{1}{2}\widehat{L}_n^\dagger\widehat{L}_n\}\rangle-\langle \widehat{L}_n^\dagger \widehat{A}_\mathrm{a} \widehat{L}_n\rangle\right)\\
    &=-\hbar\sum_{\mathbf{k},\mathbf{k}',n}(\langle\{\widehat{A}_{11},\widehat{L}_n^\dagger\widehat{L}_n\}\rangle+\mathrm{Re}\langle\{\widehat{A}_{12},\widehat{L}_n^\dagger\widehat{L}_n\}\rangle-2\langle\widehat{L}_n^\dagger\widehat{A}_{11}\widehat{L}_n\rangle-2\mathrm{Re}\langle\widehat{L}_n^\dagger\widehat{A}_{12}\widehat{L}_n\rangle).
\end{aligned}
\label{ALLD2}
\end{equation}
Consequently, we have
\begin{equation}
\begin{aligned}
    &\frac{1}{2}\langle\{\widehat{A}_{11},\widehat{L}_n^\dagger\widehat{L}_n\}\rangle-\langle\widehat{L}_n^\dagger\widehat{A}_{11}\widehat{L}_n\rangle=-2\gamma a_{\mathbf{k},\mathbf{k}'}\rho_{\mathbf{k}',\mathbf{k}}\\
    &+\frac{a_{\mathbf{k},\mathbf{k}'} g_I}{V}\sum_{\mathbf{P},\mathbf{q},\mathbf{q}'}\delta_{\mathbf{k}',\frac{\mathbf{P}}{2}-\mathbf{q}}\rho_{\frac{\mathbf{P}}{2}-\mathbf{q}',\mathbf{k}}\rho_{\frac{\mathbf{P}}{2}+\mathbf{q}',\frac{\mathbf{P}}{2}+\mathbf{q}}+\frac{a_{\mathbf{k},\mathbf{k}'} g_I}{V}\sum_{\mathbf{P},\mathbf{q},\mathbf{q}'}\delta_{\mathbf{k}',\frac{\mathbf{P}}{2}-\mathbf{q}}\kappa^*_{\mathbf{k},\frac{\mathbf{P}}{2}+\mathbf{q}}\kappa_{\frac{\mathbf{P}}{2}-\mathbf{q}',\frac{\mathbf{P}}{2}+\mathbf{q}'}\\
    &+\frac{a_{\mathbf{k},\mathbf{k}'} g_I}{V}\sum_{\mathbf{P},\mathbf{q},\mathbf{q}'}\delta_{\mathbf{k},\frac{\mathbf{P}}{2}-\mathbf{q}'}\rho_{\mathbf{k}',\frac{\mathbf{P}}{2}-\mathbf{q}}\rho_{\frac{\mathbf{P}}{2}+\mathbf{q}',\frac{\mathbf{P}}{2}+\mathbf{q}}+\frac{a_{\mathbf{k},\mathbf{k}'} g_I}{V}\sum_{\mathbf{P},\mathbf{q},\mathbf{q}'}\delta_{\mathbf{k},\frac{\mathbf{P}}{2}-\mathbf{q}'}\kappa^*_{\frac{\mathbf{P}}{2}-\mathbf{q},\frac{\mathbf{P}}{2}+\mathbf{q}}\kappa_{\mathbf{k}',\frac{\mathbf{P}}{2}+\mathbf{q}'},
\end{aligned}
\label{AL11}
\end{equation} 
and
\begin{equation}
\begin{aligned}
    &\frac{1}{2}\langle\{\widehat{A}_{12},\widehat{L}_n^\dagger\widehat{L}_n\}\rangle-\langle\widehat{L}_n^\dagger\widehat{A}_{12}\widehat{L}_n\rangle=\frac{g_Ib_{\mathbf{k},\mathbf{k}'}^*}{V}\sum_{\mathbf{P},\mathbf{q},\mathbf{q}'}\delta_{\mathbf{k},\frac{\mathbf{P}}{2}-\mathbf{q}}\delta_{\mathbf{k}',\frac{\mathbf{P}}{2}+\mathbf{q}}\kappa_{\frac{\mathbf{P}}{2}-\mathbf{q}',\frac{\mathbf{P}}{2}+\mathbf{q}'}\\
    &-\frac{b_{\mathbf{k},\mathbf{k}'}^* g_I}{V}\sum_{\mathbf{P},\mathbf{q},\mathbf{q}'}\delta_{\mathbf{k}',\frac{\mathbf{P}}{2}+\mathbf{q}}\rho_{\frac{\mathbf{P}}{2}-\mathbf{q}',\frac{\mathbf{P}}{2}-\mathbf{q}}\kappa_{\frac{\mathbf{P}}{2}+\mathbf{q}',\mathbf{k}}-\frac{b_{\mathbf{k},\mathbf{k}'}^* g_I}{V}\sum_{\mathbf{P},\mathbf{q},\mathbf{q}'}\delta_{\mathbf{k}',\frac{\mathbf{P}}{2}+\mathbf{q}}\rho_{\mathbf{k},\frac{\mathbf{P}}{2}-\mathbf{q}}\kappa_{\frac{\mathbf{P}}{2}-\mathbf{q}',\frac{\mathbf{P}}{2}+\mathbf{q}'}\\
    &-\frac{b_{\mathbf{k},\mathbf{k}'}^* g_I}{V}\sum_{\mathbf{P},\mathbf{q},\mathbf{q}'}\delta_{\mathbf{k},\frac{\mathbf{P}}{2}-\mathbf{q}}\rho_{\frac{\mathbf{P}}{2}+\mathbf{q}',\frac{\mathbf{P}}{2}+\mathbf{q}}\kappa_{\frac{\mathbf{P}}{2}-\mathbf{q}',\mathbf{k}'}-\frac{b_{\mathbf{k},\mathbf{k}'}^* g_I}{V}\sum_{\mathbf{P},\mathbf{q},\mathbf{q}'}\delta_{\mathbf{k},\frac{\mathbf{P}}{2}-\mathbf{q}}\rho_{\mathbf{k}',\frac{\mathbf{P}}{2}+\mathbf{q}}\kappa_{\frac{\mathbf{P}}{2}-\mathbf{q}',\frac{\mathbf{P}}{2}+\mathbf{q}'}.
\end{aligned}
\label{AL12}
\end{equation}
By applying Wick's theorem to the energy functional $\operatorname{Tr}(\widehat{D}\widehat{H}_{I})\equiv\langle\widehat{H}_{I}\rangle$, we have
\begin{equation}
    \begin{aligned}
        \langle H_I\rangle&=-\gamma \sum_\mathbf{q}(\rho_{\uparrow,\mathbf{q},\mathbf{q}}+\rho_{\downarrow,\mathbf{q},\mathbf{q}})\\
        &+\frac{g_I}{V}\sum_{\mathbf{P},\mathbf{q},\mathbf{q}'}\rho_{\uparrow,\frac{\mathbf{P}}{2}-\mathbf{q},\frac{\mathbf{P}}{2}-\mathbf{q}'}\rho_{\downarrow,\frac{\mathbf{P}}{2}+\mathbf{q},\frac{\mathbf{P}}{2}+\mathbf{q}'}+\frac{g_I}{V}\sum_{\mathbf{P},\mathbf{q},\mathbf{q}'}\kappa^*_{\uparrow,\frac{\mathbf{P}}{2}-\mathbf{q}',\frac{\mathbf{P}}{2}+\mathbf{q}'}\kappa_{\uparrow,\frac{\mathbf{P}}{2}-\mathbf{q},\frac{\mathbf{P}}{2}+\mathbf{q}},
    \end{aligned}
    \end{equation}
from which we know the form of $h_I$ and $\Delta_I$:
\begin{align}
    &(h_I)_{\mathbf{k},\mathbf{k}'}=-\gamma \delta_{\mathbf{k},\mathbf{k}'}+\frac{g_I}{V}\sum_{\mathbf{P},\mathbf{q},\mathbf{q}'}\delta_{\frac{\mathbf{P}}{2}-\mathbf{q}',\mathbf{k}'}\delta_{\frac{\mathbf{P}}{2}-\mathbf{q},\mathbf{k}}\rho_{\frac{\mathbf{P}}{2}+\mathbf{q}',\frac{\mathbf{P}}{2}+\mathbf{q}},\label{hI}\\
    &(\Delta_I)_{\mathbf{k},\mathbf{k}'}=\frac{g_I}{V}\sum_{\mathbf{P},\mathbf{q},\mathbf{q}'}\delta_{\frac{\mathbf{P}}{2}-\mathbf{q},\mathbf{k}}\delta_{\frac{\mathbf{P}}{2}+\mathbf{q},\mathbf{k}'}\kappa_{\frac{\mathbf{P}}{2}-\mathbf{q}',\frac{\mathbf{P}}{2}+\mathbf{q}'}.\label{DeltaI}
\end{align}
From Eqs.~(\ref{AL11}), (\ref{AL12}), (\ref{hI}) and (\ref{DeltaI}), we can arrange Eq.~(\ref{ALLD2}) into
\begin{equation}
    -\hbar\Tr \widehat{A}_\mathrm{a}\sum_n\left(\frac{1}{2}\{\widehat{L}^\dagger_n\widehat{L}_n,\widehat{D}_\mathrm{a}\}-\widehat{L}_n\widehat{D}_\mathrm{a}\widehat{L}^\dagger_n\right)=-\hbar(T_1+T_2),
\end{equation}
where
\begin{equation}
\begin{aligned}
    T_1=&\sum_{\mathbf{k}}\delta_{\mathbf{k},\mathbf{k}'}\sum_{\mathbf{q},\mathbf{q}'}a_{\mathbf{k},\mathbf{q}}(h_I)_{\mathbf{q},\mathbf{q}'}\rho_{\mathbf{q}',\mathbf{k}'}+a_{\mathbf{k},\mathbf{q}}(\Delta_I)_{\mathbf{q},\mathbf{q}'}\kappa^\dagger_{\mathbf{q}',\mathbf{k}'}+a_{\mathbf{k},\mathbf{q}}\kappa_{\mathbf{q},\mathbf{q}'}(\Delta_I^\dagger)_{\mathbf{q}',\mathbf{k}'}+a_{\mathbf{k},\mathbf{q}}\rho_{\mathbf{q},\mathbf{q}'}(h_I)_{\mathbf{q}',\mathbf{k}'}\\
    &+b_{\mathbf{k},\mathbf{q}}(\Delta_I^\dagger)_{\mathbf{q},\mathbf{q}'}\rho_{\mathbf{q}',\mathbf{k}'}+b_{\mathbf{k},\mathbf{q}}\kappa^\dagger_{\mathbf{q},\mathbf{q}'}(h_I)_{\mathbf{q}',\mathbf{k}'}-b_{\mathbf{k},\mathbf{q}}(h_I^*)_{\mathbf{q},\mathbf{q}'}\kappa^\dagger_{\mathbf{q}',\mathbf{k}'}+b_{\mathbf{k},\mathbf{q}}(\delta_{\mathbf{q},\mathbf{q}'}-\rho^*_{\mathbf{q},\mathbf{q}'})(\Delta_I^\dagger)_{\mathbf{q}',\mathbf{k}'}\\
    &+b^\dagger_{\mathbf{k},\mathbf{q}}(h_I)_{\mathbf{q},\mathbf{q}'}\kappa_{\mathbf{q}',\mathbf{k}'}+b^\dagger_{\mathbf{k},\mathbf{q}}(\Delta_I)_{\mathbf{q},\mathbf{q}'}(\delta_{\mathbf{q}',\mathbf{k}'}-\rho^*_{\mathbf{q}',\mathbf{k}'})-b^\dagger_{\mathbf{k},\mathbf{q}}\kappa_{\mathbf{q},\mathbf{q}'}(h_I^*)_{\mathbf{q}',\mathbf{k}'}+b^\dagger_{\mathbf{k},\mathbf{q}}\rho_{\mathbf{q},\mathbf{q}'}(\Delta_I)_{\mathbf{q}',\mathbf{k}'}\\
    &-a^*_{\mathbf{k},\mathbf{q}}(\Delta_I^\dagger)_{\mathbf{q},\mathbf{q}'}\kappa_{\mathbf{q}',\mathbf{k}'}-a^*_{\mathbf{k},\mathbf{q}}\kappa^\dagger_{\mathbf{q},\mathbf{q}'}(\Delta_I)_{\mathbf{q}',\mathbf{k}'}+a^*_{\mathbf{k},\mathbf{q}}(h_I^*)_{\mathbf{q},\mathbf{q}'}(\delta_{\mathbf{q}',\mathbf{k}'}-\rho^*_{\mathbf{q}',\mathbf{k}'})+a^*_{\mathbf{k},\mathbf{q}}(\delta_{\mathbf{q},\mathbf{q}'}-\rho^*_{\mathbf{q},\mathbf{q}'})(h_I^*)_{\mathbf{q}',\mathbf{k}'}\\
    =&\tr\mathcal{A}\{\mathcal{H}_I,\mathcal{R}\},
\end{aligned}
\end{equation}
and
\begin{equation}
    \begin{aligned}
        T_2=&\sum_{\mathbf{k}}\delta_{\mathbf{k},\mathbf{k}'}\sum_{\mathbf{q},\mathbf{q}'}2b_{\mathbf{k},\mathbf{q}}(h_I^*)_{\mathbf{q},\mathbf{q}'}\kappa^\dagger_{\mathbf{q}',\mathbf{k}'}-2b_{\mathbf{k},\mathbf{q}}(\Delta_I^\dagger)_{\mathbf{q},\mathbf{q}'}\rho_{\mathbf{q}',\mathbf{k}'}\\
        &+2b^\dagger_{\mathbf{k},\mathbf{q}}\kappa_{\mathbf{q},\mathbf{q}'}(h_I^*)_{\mathbf{q}',\mathbf{k}'}-2b^\dagger_{\mathbf{k},\mathbf{q}}\rho_{\mathbf{q},\mathbf{q}'}(\Delta_I)_{\mathbf{q}',\mathbf{k}'}\\
        &-a^*_{\mathbf{k},\mathbf{q}}(h_I^*)_{\mathbf{q},\mathbf{q}'}(\delta_{\mathbf{q}',\mathbf{k}'}-2\rho^*_{\mathbf{q}',\mathbf{k}'})-a^*_{\mathbf{k},\mathbf{q}}(\delta_{\mathbf{q},\mathbf{q}'}-2\rho^*_{\mathbf{q},\mathbf{q}'})(h_I^*)_{\mathbf{q}',\mathbf{k}'}\\
        &+2a^*_{\mathbf{k},\mathbf{q}}(\Delta_I^\dagger)_{\mathbf{q},\mathbf{q}'}\kappa_{\mathbf{q}',\mathbf{k}'}+2a^*_{\mathbf{k},\mathbf{q}}\kappa^\dagger_{\mathbf{q},\mathbf{q}'}(\Delta_I)_{\mathbf{q}',\mathbf{k}'}\\
        =&\tr\mathcal{A}\mathcal{J}.
\end{aligned}
\end{equation}
Combining all the above results, we reach the effective action Eq.~(\ref{Seff}) in the main text. For further derivation, it is convenient to separate the equations of motion for four sub-blocks of Eq.~(\ref{TDHFB}), which are
\begin{align}
    &i\hbar\dot{\rho}=[h_R,\rho]+i\{h_I,\rho\}+\Delta_R\kappa^\dagger-\kappa\Delta^\dagger_R+i\kappa\Delta_I^\dagger+i\Delta_I\kappa^\dagger,\label{TDHFB1}\\
    &i\hbar\dot{\kappa}=h_R\kappa+\kappa h_R^*+ih_I\kappa+i\kappa h_I^*+\Delta_R(1-\rho^*)-\rho\Delta_R+i\Delta_I(1-\rho^*)-i\rho\Delta_I,\label{TDHFB2}\\
    &i\hbar\dot{\kappa}^\dagger=-h_R^*\kappa^\dagger-\kappa^\dagger h_R+ih_I^*\kappa^\dagger+i\kappa^\dagger h_I-(1-\rho^*)\Delta^\dagger+\Delta_R^\dagger \rho+i(1-\rho^*)\Delta_I^\dagger-i\Delta_I^\dagger\rho,\label{TDHFB3}\\
    &-i\hbar\dot{\rho}^*=-[h_R^*,\rho^*]+i\{h_I^*,\rho^*\}-\Delta_R^\dagger\kappa+\kappa^\dagger\Delta_R+i\kappa^\dagger\Delta_I+i\Delta_I^\dagger\kappa.\label{TDHFB4}
\end{align}

\section{Detailed Calculation of Semi-Classic Approximation}

In this section, we provide a detailed calculation to obtain Eqs.~(\ref{nu_IQBE}) and (\ref{continuity_eqn}) in the main text from Eqs.~(\ref{TDHFB1}), (\ref{TDHFB2}), (\ref{TDHFB3}), and (\ref{TDHFB4}) derived in the last section. 

Because Eqs.~(\ref{TDHFB1}), (\ref{TDHFB2}), (\ref{TDHFB3}), and (\ref{TDHFB4}) are matrix equations that are independent of basis, we first choose them to be represented in the real space. This allows us to change complex conjugation operations to time reversal operations (denoted by a bar hat $\bar{\cdot}$) due to their equivalence in the real space. 

Then we perform the Wigner transform to all terms in Eqs.~(\ref{TDHFB1}), (\ref{TDHFB2}), (\ref{TDHFB3}), and (\ref{TDHFB4}). The Wigner transform for an arbitrary matrix $O$ is defined by
\begin{equation}
O(\mathbf{r}, \mathbf{p})=\int d^3 s e^{-i \mathbf{p} \cdot \mathbf{s} / \hbar} O_{\mathbf{r}+\frac{\mathbf{s}}{2}, \mathbf{r}-\frac{\mathbf{s}}{2}}.
\end{equation} 
When the Wigner transform is performed, there are two important identities:
\begin{equation}
\bar{O}(\mathbf{r}, \mathbf{p})=O(\mathbf{r},-\mathbf{p}), \quad\left[O^{\dagger}\right](\mathbf{r}, \mathbf{p})=O^*(\mathbf{r}, \mathbf{p}).
\end{equation}
Besides, it is worth noting that the Wigner transform turns all multiplications to Moyal products denoted by a $\star$. For our aim, it is enough to expand the Moyal product to the first order in $\hbar$:
\begin{align*}
    A(\mathbf{r},\mathbf{p})\star B(\mathbf{r},\mathbf{p})=AB+\frac{i\hbar}{2}\{A,B\}_{c}+\mathcal{O}(\hbar^2),
\end{align*}
where $\{,\}_c$ is Poisson bracket, i.e.,
\begin{equation}
    \{A,B\}_c \equiv \sum_{i=x, y, z}\left(\frac{\partial A}{\partial r_i} \frac{\partial B}{\partial p_i}-\frac{\partial A}{\partial p_i} \frac{\partial B}{\partial r_i}\right).
\end{equation}
We rewrite Eqs.~(\ref{TDHFB1}), (\ref{TDHFB2}), (\ref{TDHFB3}), and (\ref{TDHFB4}) in the phase space to be (For convenience, we do not write the arguments $(\mathbf{r},\mathbf{p})$ of functions below. All multiplications below are scalar multiplication rather than matrix product.)
\begin{equation}
i\hbar\dot{\rho}=i\hbar\{h_R,\rho\}_c-2i\mathrm{Im}(\Delta_R^*\kappa)+i\hbar\mathrm{Re}\{\Delta_R^*,\kappa\}_c+2i\hbar h_I\rho+2i\hbar\mathrm{Re}(\Delta_I^*\kappa)+i\hbar\mathrm{Im}\{\Delta_R^*,\kappa\}_c,
\end{equation}
\begin{equation}
\begin{aligned}
i\hbar\dot{\kappa}=&(h_R+\bar{h}_R)\kappa+\frac{i\hbar}{2}\{h_R-\bar{h}_R,\kappa\}_c+[1-(\rho+\bar{\rho})]\Delta_R+\frac{i\hbar}{2}\{\Delta_R,\rho-\bar{\rho}\}_c\\
&+i(h_I+\bar{h}_I)\kappa-\frac{\hbar}{2}\{h_I-\bar{h}_I,\kappa\}_c+i[1-(\rho+\bar{\rho})]\Delta_I-\frac{\hbar}{2}\{\Delta_I,\rho-\bar{\rho}\}_c,
\end{aligned}
\end{equation}
\begin{equation}
\begin{aligned}
i\hbar\dot{\kappa}^*=&-(h_R+\bar{h}_R)\kappa^*+\frac{i\hbar}{2}\{h_R-\bar{h}_R,\kappa^*\}_c-[1-(\rho+\bar{\rho})]\Delta_R^*+\frac{i\hbar}{2}\{\Delta_R^*,\rho-\bar{\rho}\}_c\\
&+i(h_I+\bar{h}_I)\kappa^*+\frac{\hbar}{2}\{h_I-\bar{h}_I,\kappa^*\}_c+i[1-(\rho+\bar{\rho})]\Delta_I^*+\frac{\hbar}{2}\{\Delta_I^*,\rho-\bar{\rho}\}_c,
\end{aligned}
\end{equation}
\begin{equation}
i\hbar\dot{\bar{\rho}}=-i\hbar\{\bar{h}_R,\bar{\rho}\}_c-2i\mathrm{Im}(\Delta_R^*\kappa)-i\hbar\mathrm{Re}\{\Delta_R^*,\kappa\}_c+2i\hbar \bar{h}_I\bar{\rho}+2i\hbar\mathrm{Re}(\Delta_I^*\kappa)-i\hbar\mathrm{Im}\{\Delta_R^*,\kappa\}_c.
\end{equation}
Based on the expression of $h_I$ given in the main text, we note that $h_I=\bar{h}_I$. Besides, we choose to work in a gauge that ensures $\mathrm{Im}(\Delta_R)=\mathrm{Im}(\Delta_I)=0$. Those conditions simplify the set of equations to be
\begin{equation}
i\hbar\dot{\rho}=i\hbar\{h_R,\rho\}_c-2i\Delta_R\mathrm{Im}(\kappa)+i\hbar\{\Delta_R,\mathrm{Re}(\kappa)\}_c+2i\hbar h_I\rho+2i\hbar\Delta_I\mathrm{Re}(\kappa)+i\hbar\{\Delta_R,\mathrm{Im}(\kappa)\}_c,
\label{rho_eom}
\end{equation}
\begin{equation}
\begin{aligned}
i\hbar\dot{\kappa}=&(h_R+\bar{h}_R)\kappa+\frac{i\hbar}{2}\{h_R-\bar{h}_R,\kappa\}_c+[1-(\rho+\bar{\rho})]\Delta_R+\frac{i\hbar}{2}\{\Delta_R,\rho-\bar{\rho}\}_c\\
&+2ih_I\kappa+i[1-(\rho+\bar{\rho})]\Delta_I-\frac{\hbar}{2}\{\Delta_I,\rho-\bar{\rho}\}_c,
\end{aligned}
\label{kappa_eom}
\end{equation}
\begin{equation}
\begin{aligned}
i\hbar\dot{\kappa}^*=&-(h_R+\bar{h}_R)\kappa^*+\frac{i\hbar}{2}\{h_R-\bar{h}_R,\kappa^*\}_c-[1-(\rho+\bar{\rho})]\Delta_R^*+\frac{i\hbar}{2}\{\Delta_R^*,\rho-\bar{\rho}\}_c\\
&+2ih_I\kappa^*+i[1-(\rho+\bar{\rho})]\Delta_I^*+\frac{\hbar}{2}\{\Delta_I^*,\rho-\bar{\rho}\}_c,
\end{aligned}
\label{kappastar_eom}
\end{equation}
\begin{equation}
i\hbar\dot{\bar{\rho}}=-i\hbar\{\bar{h}_R,\bar{\rho}\}_c-2i\Delta_R\mathrm{Im}(\kappa)-i\hbar\{\Delta_R,\mathrm{Re}(\kappa)\}_c+2i\hbar \bar{h}_I\bar{\rho}+2i\hbar\Delta_I\mathrm{Re}(\kappa)-i\hbar\{\Delta_R,\mathrm{Im}(\kappa)\}_c.
\label{rhobar_eom}
\end{equation}
In the above equations, terms of different orders in $\hbar$ are mixed. Given that we only expand the Moyal product to the first order, we should only consider terms of order $\mathcal{O}(\hbar)$. Recalling that we express $H$ as $H = H_R + i\hbar H_I$, consistency requires that we also decompose $\kappa$ as $\kappa_R + i\hbar \kappa_I$, i.e., $\kappa_R=\mathrm{Re}(\kappa)=(\kappa+\kappa^*)/2$ and $\kappa_I=\mathrm{Im}(\kappa)/\hbar=(\kappa-\kappa^*)/(2i\hbar)$. Analogously, we also define the time-even and time-odd parts of $\rho$ to be $\rho^\mathrm{ev}=(\rho+\bar{\rho})/2$ and $\rho^\mathrm{od}=(\rho-\bar{\rho})/2$. Similar time-even and time-odd components can also be defined for $h_R$.  Finally, the equations of motion for $\kappa_R$, $\kappa_I$, $\rho^\mathrm{ev}$ and $\rho^\mathrm{od}$ from Eqs.~(\ref{rho_eom}), (\ref{kappa_eom}), (\ref{kappastar_eom}), and (\ref{rhobar_eom}) reads,
\begin{align}
&\hbar\dot{\rho}^{\mathrm{ev}}=\hbar\{h_R^\mathrm{ev},\rho^\mathrm{od}\}_c+\hbar\{h_R^\mathrm{od},\rho^\mathrm{ev}\}_c-2\hbar\Delta_R\kappa_I+2\hbar h_I \rho^\mathrm{ev}+2\hbar\Delta_I \kappa_R,\\
&\hbar\dot{\rho}^{\mathrm{od}}=\hbar\{h_R^\mathrm{ev},\rho^\mathrm{ev}\}_c+\hbar\{h_R^\mathrm{od},\rho^\mathrm{od}\}_c+\hbar\{\Delta_R,\kappa_R\}_c+2\hbar h_I \rho^\mathrm{od}+\hbar^2\{\Delta_I,\kappa_I\}_c,\\
&\hbar\dot{\kappa}_R=2\hbar h_R^\mathrm{ev}\kappa_I+\hbar\{h_R^\mathrm{od},\kappa_R\}_c+\hbar\{\Delta_R,\rho^\mathrm{od}\}_c+2\hbar h_I \kappa_R+\hbar(1-2\rho^\mathrm{ev})\Delta_I,\\
&\hbar^2\dot{\kappa}_I=-2h_R^\mathrm{ev}\kappa_R+\hbar^2\{h_R^\mathrm{od},\kappa_I\}_c-(1-2\rho^\mathrm{ev})\Delta_R+2\hbar^2h_I\kappa_I+\hbar^2\{\Delta_I,\rho^\mathrm{od}\}_c.
\end{align}
Ignoring terms with order $\mathcal{O}(\hbar^2)$, we obtain
\begin{align}
&\dot{\rho}^{\mathrm{ev}}=\{h_R^\mathrm{ev},\rho^\mathrm{od}\}_c+\{h_R^\mathrm{od},\rho^\mathrm{ev}\}_c-2\Delta_R\kappa_I+2h_I \rho^\mathrm{ev}+2\Delta_I \kappa_R,\label{rhoev_eom}\\
&\dot{\rho}^{\mathrm{od}}=\{h_R^\mathrm{ev},\rho^\mathrm{ev}\}_c+\{h_R^\mathrm{od},\rho^\mathrm{od}\}_c+\{\Delta_R,\kappa_R\}_c+2 h_I \rho^\mathrm{od},\label{rhood_eom}\\
&\dot{\kappa}_R=2 h_R^\mathrm{ev}\kappa_I+\{h_R^\mathrm{od},\kappa_R\}_c+\{\Delta_R,\rho^\mathrm{od}\}_c+2 h_I \kappa_R+(1-2\rho^\mathrm{ev})\Delta_I,\label{kappaR_eom}\\
&(1-2\rho^\mathrm{ev})\Delta_R=-2h_R^\mathrm{ev}\kappa_R.\label{kappaI_eom}
\end{align}
We observe that Eq.~(\ref{kappaI_eom}) becomes an identity, indicating a well-defined Bogoliubov quasi-particle as in the conventional BCS theory. 
If one can not ignore $\mathcal{O}(\hbar^2)$ terms here, e.g., for strongly interacting systems, the quasiparticle will have a finite lifetime, rendering the dynamics of quasiparticle distribution function difficult to calculate. At the same time, we mention that in the corresponding Hermitian theory, the second-order term is crucial for obtaining a soliton solution of oscillating order parameter under interaction quench.
We define the time-even quasi-particle distribution function $\nu^\mathrm{ev}$ to be
\begin{equation}
\frac{h_R^\mathrm{ev}\Delta_R(1-2\nu^\mathrm{ev})}{E^\mathrm{ev}}=(1-2\rho^\mathrm{ev})\Delta_R=-2h_R^\mathrm{ev}\kappa_R,
\end{equation}
where the time-even quasi-particle energy has a gapped dispersion relation
\begin{equation}
E^\mathrm{ev}=\sqrt{\left(h_R^{\mathrm{ev}}\right)^2+\Delta_R^2}.
\end{equation}
The advantage of defining a quasi-particle distribution is that we can use it to hide $\kappa_I$ appearing in Eqs.~(\ref{rhoev_eom}) and (\ref{kappaR_eom}). Multiplying two equations with $h_R^\mathrm{ev}$ and $\Delta_R$ and adding them up, we have
\begin{equation}
\dot{\nu}^\mathrm{ev}=\{E^\mathrm{ev},\rho^\mathrm{od}\}_c+\{h_R^\mathrm{od},\nu^\mathrm{ev}\}_c+h_I\left(2\nu^\mathrm{ev}-1+\frac{h_R^\mathrm{ev}}{E^\mathrm{ev}}\right).\label{nuev_eom}
\end{equation}
Equation~(\ref{rhood_eom}) is well-behaved without any dependence on $\kappa_I$, which means the time-odd and time-even motions are entirely decoupled. Consequently, we can directly define the time-odd quasi-particle distribution to be the time-odd physical particle distribution itself, and the quasi-particle energy is nothing but the original time-odd Hartree-Fock Hamiltonian:
\begin{equation}
\nu^\mathrm{od}=\rho^\mathrm{od},\quad E^\mathrm{od}=h_R^\mathrm{od}.
\end{equation}
Then, by combining Eqs.~(\ref{rhood_eom}) and (\ref{nuev_eom}), we derive the equation of motion of the complete quasi-particle distribution $\nu=\nu^\mathrm{ev}+\nu^\mathrm{od}$,
\begin{equation}
	\dot{\nu}=\{E,\nu\}_c+h_I\left(2\nu-1+\frac{h_R^\mathrm{ev}}{E^\mathrm{ev}}\right),
\end{equation}
where the total quasi-particle energy $E=E^\mathrm{ev}+E^\mathrm{od}$. 

Equation~(\ref{nuev_eom}) by itself does not close because we do not explicitly know the gauge $\phi(\mathbf{r})$ in $h_R$. To determine $\phi(\mathbf{r})$, one should recall our requirement of choosing the gauge $\mathrm{Im}(\Delta_{R/I})=0$. Expressing $\Delta_{R/I}$ using the gap equation Eq.~(\ref{gap_eqn}) in the main text,
\begin{equation}
    \int \frac{d^3 p}{(2 \pi \hbar)^3}\kappa_I(\mathbf{r}, \mathbf{p})=0.
\end{equation}
Substituting to Eq.~(\ref{rhoev_eom}), we obtain
\begin{equation}
    \int d^3p [\dot{\rho}^{\mathrm{ev}}-\{h_R^\mathrm{ev},\rho^\mathrm{od}\}_c-\{h_R^\mathrm{od},
    \rho^\mathrm{ev}\}_c-2h_I \rho^\mathrm{ev}-2\Delta_I \kappa_R]=0.
    \label{rhoev_eom_2}
\end{equation}

\section{Calculation on Dynamics of Superfluid Fraction and Superconducting Gap in Homogeneous Systems}

\subsection{Superfluid Fraction}

We rearrange Eq.~(\ref{superfluid_fraction}) into a dimensionless form:
\begin{equation}
    \zeta(t)=1+\frac{n(0)}{n(t)}\int_0^\infty d\bar{p} \bar{p}^3 \frac{\partial \bar{\nu}}{\partial \bar{p}}\frac{\sqrt{\bar{\Delta}_R^2+(\bar{p}^2-\bar{\mu})^2}}{\bar{p}^2-\bar{\mu}}\xrightarrow[]{\Delta_R\rightarrow 0^+}1+\frac{n(0)}{n(t)}\int_0^\infty d\bar{p} \bar{p}^3 \frac{\partial \bar{\nu}}{\partial \bar{p}}\operatorname{sgn}(\bar{p}^2-\bar{\mu}),
\end{equation}
by defining $\bar{p}=p/p_F$, $\bar{\nu}=\nu p_F$, $\bar{\rho}=\rho p_F$, $\bar{\Delta}_R=\Delta_R/E_F$ and $\bar{\mu}=(\mu-\dot{\phi})/E_F$ with $E_F=p_F^2/2m$.
As $\nu$ is also solved in the BCS limit in this work to be $\bar{\nu}=\alpha(t)\theta(1-\bar{p})$, it is easy to note that directly evaluating $\zeta$ using the above expression will yield an ill-defined result. To resolve the problem, we note from Eq.~(\ref{rhoev_to_nu}) that when $\Delta_R\rightarrow0^+$, (similar to the main text, we ignore superscript $(\cdot)^\mathrm{ev}$)
\begin{equation}
    \bar{\nu}(\bar{p})=1-\bar{\rho}(\bar{p})\quad \text{if $\bar{p}<1$},
\end{equation}
and
\begin{equation}
    \operatorname{sgn}(\bar{p}^2-\bar{\mu})=\operatorname{sgn}[1-2\bar{\rho}(\bar{p})].
    \label{xy_equivalence}
\end{equation}
Therefore,
\begin{equation}
\begin{aligned}
    \zeta(t)&=1-\frac{n(0)}{n(t)}\int_0^\infty d\bar{p} \bar{p}^3 \frac{\partial \bar{\rho}}{\partial \bar{p}}\operatorname{sgn}(1-2\bar{\rho}(\bar{p}))\\
    &=1-\frac{n(0)}{n(t)}\int_0^\infty d\bar{p} \bar{p}^3 \frac{\partial \bar{\rho}_1}{\partial \bar{p}}+\frac{n(0)}{n(t)}\int_0^\infty d\bar{p} \bar{p}^3 \frac{\partial \bar{\rho}_2}{\partial \bar{p}},
\end{aligned}
\end{equation}
where
\begin{equation}
    \bar{\rho}_1=[1-\alpha(t)]\theta(1-\bar{p})+\frac{1}{2}\theta(\bar{p}-1)\ge\frac{1}{2},\quad \bar{\rho}_2=-\frac{1}{2}\theta(\bar{p}-1)\le 0,\quad \bar{\rho}=\bar{\rho}_1+\bar{\rho}_2.
    \label{rho_decomposition}
\end{equation}
Then, it is straightforward to obtain
\begin{equation}
\begin{aligned}
    \zeta(t)&=1-\frac{n(0)}{n(t)}\int_0^\infty d\bar{p} \bar{p}^3 (\alpha-\frac{1}{2})\delta(\bar{p}-1)-\frac{n(0)}{n(t)}\int_0^\infty d\bar{p} \bar{p}^3 \frac{1}{2}\delta(\bar{p}-1)\\
    &=1-\frac{n(0)}{n(t)}\alpha(t)=\frac{1-2\alpha(t)}{1-\alpha(t)}.
\end{aligned}
\end{equation}

\subsection{Superfluid Gap}

The gap equation for $\Delta_R$ [Eq.~(\ref{gap_eqn}) in the main text] in the dimensionless form reads
\begin{equation}
    1=-\frac{2\mathrm{Re}(\bar{a}_s)}{\pi}\int_0^\infty d\bar{p} \bar{p}^2\left[\frac{[1-2\bar{\nu}(\bar{p})]}{\sqrt{(\bar{p}^2-\bar{\mu})^2+\bar{\Delta}_R^2}}-\frac{1}{\bar{p}^2}\right],
\end{equation}
where $\mathrm{Re}(\bar{a}_s)\equiv \mathrm{Re}(a_s)p_F/\hbar$. 
By changing variable from $\bar{p}^2$ to $x$, we obtain
\begin{equation}
    -\frac{\pi}{\mathrm{Re}(\bar{a}_s)}=\mathcal{P}\int_0^\infty dx x^{1/2}\left[\frac{1-2\bar{\nu}(\sqrt{x})}{\sqrt{(x-\bar{\mu})^2+\bar{\Delta}_R^2}}-\frac{1}{x}\right],
\end{equation}
where, similar to the calculation in the standard BCS theory, we interpret the integral by its Cauchy principal value. If $\bar{\mu}>0$, we have
\begin{equation}
    \mathcal{P}\int_0^\infty dx x^{1/2}\left(\frac{1}{x}-\frac{1}{x-\bar{\mu}}\right)=0.
\end{equation}
Then, we rewrite the right-hand side of the gap equation into three terms,
\begin{equation}
\begin{aligned}
    -\frac{\pi}{\mathrm{Re}(\bar{a}_s)}&=\mathcal{P}\int_0^\infty dx (x^{1/2}-1)\left[\frac{1}{\sqrt{(x-\bar{\mu})^2+\bar{\Delta}_R^2}}-\frac{1}{x-\bar{\mu}}\right]\\
    &+\mathcal{P} \int_0^\infty dx \left[\frac{1}{\sqrt{(x-\bar{\mu})^2+\bar{\Delta}_R^2}}-\frac{1}{x-\bar{\mu}}\right]\\
    &-\int_0^\infty dx \left[\frac{2x^{1/2}\bar{\nu}(\sqrt{x})}{\sqrt{(x-\bar{\mu})^2+\bar{\Delta}_R^2}}\right].
\end{aligned}
\end{equation}
The first two terms simplify under the limit $\Delta_R\rightarrow0$,
\begin{equation}
    -\frac{\pi}{\mathrm{Re}(\bar{a}_s)}=2\sqrt{\bar{\mu}}[\ln(4)-2]+2\ln(2\bar{\mu})-2\ln(\bar{\Delta}_R)-\int_0^\infty dx \left[\frac{2x^{1/2}\bar{\nu}(\sqrt{x})}{\sqrt{(x-\bar{\mu})^2+\bar{\Delta}_R^2}}\right].
    \label{gap_eqn_simplified}
\end{equation}
From Eqs.~(\ref{xy_equivalence}) and (\ref{rho_decomposition}), we find $\bar{\mu}=1$ is a constant. Thus, expanding Eq.~(\ref{gap_eqn_simplified}) to the first order of $\bar{\Delta}_R$ and solving $\bar{\Delta}_R$, we have
\begin{equation}
    \bar{\Delta}_R=\frac{8}{e^2}\exp\left\{\dfrac{\pi}{[2\mathrm{Re}(\bar{a}_s)(1-2\alpha(t))]}\right\}.
\end{equation}

\end{document}